# Role of Pressure in the Growth of Hexagonal Boron Nitride Thin Films from Ammonia-Borane

Justin C. Koepke[1,2,3,a)†], Joshua D. Wood[1,2,3,b)†], Yaofeng Chen[1,2,3], Scott W. Schmucker[4], Ximeng Liu[1,2,3], Noel N. Chang[5], Lea Nienhaus[2,3,5,c)], Jae Won Do[1,2,3], Enrique A. Carrion[1,3], Jayan Hewaparakrama[1,3], Aniruddh Rangarajan[1,2,3], Isha Datye[1,2,3], Rushabh Mehta[1,2,3], Richard T. Haasch[6], Martin Gruebele[2,5,7], Gregory S. Girolami[2,5], Eric Pop[1,8], and Joseph W. Lyding[1,2,3*]

[1]Dept. of Electrical & Computer Eng., Univ. of Illinois at Urbana-Champaign, Urbana, IL 61801
[2]Beckman Institute, Univ. of Illinois at Urbana-Champaign, Urbana, IL 61801
[3]Micro and Nanotechnology Lab, Univ. of Illinois at Urbana-Champaign, Urbana, IL 61801
[4]U.S. Naval Research Laboratory, Washington, DC 20375
[5]Dept. of Chemistry, Univ. of Illinois at Urbana-Champaign, Urbana, IL 61801
[6]Materials Research Laboratory, Univ. of Illinois at Urbana-Champaign, Urbana, IL 61801
[7]Dept. of Physics, Univ. of Illinois at Urbana-Champaign, Urbana, IL 61801
[8]Electrical Engineering, Stanford University, Stanford, CA 94305

a) Present address: Sandia National Laboratories, P.O. Box 5800, Albuquerque, NM 87185
b) Present address: Dept. of Materials Science and Eng., Northwestern Univ., Evanston, IL 60208
c) Present address: Dept. of Chemistry, Massachusetts Inst. of Technol., Cambridge, MA 02139

† These authors contributed equally.

We analyze the optical, chemical, and electrical properties of chemical vapor deposition (CVD) grown hexagonal boron nitride (h-BN) using the precursor ammonia-borane ($H_3N–BH_3$) as a function of $Ar/H_2$ background pressure ($P_{TOT}$). Films grown at $P_{TOT} \leq 2.0$ Torr are uniform in thickness, highly crystalline, and consist solely of h-BN. At larger $P_{TOT}$, with constant precursor flow, the growth rate increases, but the resulting h-BN is more amorphous, disordered, and $sp^3$ bonded. We attribute these changes in h-BN grown at high pressure to incomplete thermolysis of the $H_3N–BH_3$ precursor from a passivated Cu catalyst. A similar increase in h-BN growth rate and amorphization is observed even at low $P_{TOT}$ if the $H_3N–BH_3$ partial pressure is initially greater than the background pressure $P_{TOT}$ at the beginning of growth. h-BN growth using the $H_3N–BH_3$ precursor reproducibly can give large-area, crystalline h-BN thin films, provided that the total pressure is under 2.0 Torr and the precursor flux is well-controlled.

* Correspondence should be addressed to lyding@illinois.edu, jkoepkeuiuc@gmail.com, and joshua.wood@northwestern.edu.

**KEYWORDS:** hexagonal boron nitride, chemical vapor deposition (CVD), synthesis, morphology, ammonia-borane, crystallinity; monolayer

## INTRODUCTION

Hexagonal boron nitride (h-BN) is an insulating, two-dimensional (2D) equivalent of graphene. Films of h-BN have been used as insulating spacers,[1] encapsulants,[2] substrates for electronic devices,[3, 4] corrosion and oxidation-resistant coatings,[5, 6] and surfaces for growth of other 2D nanomaterials such as graphene[7] and $WS_2$.[8] Most of these studies employed small-area (~100 μm$^2$) h-BN pieces exfoliated from sintered h-BN crystals,[9] limiting technological use of h-BN films. Additionally, unlike graphene, h-BN is difficult to prepare in monolayer form by exfoliation. The electronegativity difference between B and N and the reduced resonance stabilization relative to graphene results in electrostatic attractions between layers and in-plane. Consequently, it is more challenging to control h-BN grain size and layer number. Furthermore, partially ionic B–N bonds can form between neighboring BN layers, serving to "spot weld" such layers together.[10]

Several groups have sought to overcome these limitations by using chemical vapor deposition (CVD) to grow large-area, monolayer h-BN films.[11-22] CVD growth of h-BN has been accomplished using various precursors (e.g., ammonia borane, borazine, and diborane) on transition metal substrates (e.g., Cu, Ni,[23] Fe,[24] Ru,[25, 26] etc.). Of these h-BN growth substrates, we focus on Cu, as Cu has a high catalytic activity,[27] is inexpensive, and is the typical growth substrate[28] for conventional graphene CVD. Regarding h-BN growth precursors, volatile borazine—$B_3N_3H_6$, isoelectronic with benzene—is far from an ideal choice, as borazine is hazardous and decomposes quickly even at room temperature. While borazine can pyrolyze and dehydrogenate[23, 25, 29, 30] to generate h-BN films,[13, 17, 19, 20, 22, 31] partial dehydrogenation is common,[30] resulting in oligomeric BN compounds and aperiodic h-BN grain boundaries.[13, 17] Finally, thin films of h-BN can also be grown from mixtures of diborane ($B_2H_6$) and ammonia ($NH_3$),[21] with the drawback that diborane is toxic and pyrophoric.

In contrast to borazine and diborane, the molecule ammonia-borane ($H_3N–BH_3$) is inexpensive, air stable, and has sufficient volatility to serve as a CVD precursor for h-BN thin films.[12, 15, 16, 18, 21, 32-38] High purity ammonia-borane shows no decomposition over two months at room temperature,[39] and its vapor pressure has been estimated to be ~10$^{-4}$ Torr at room temperature.[40, 41] Heating ammonia-borane generates hydrogen and volatile B- and N-containing species that enable h-BN growth; the generated species include monomeric aminoborane ($H_2N=BH_2$), borazine, and small amounts of diborane.[34, 42] However, the growth of h-BN from ammonia-borane, typically carried out in the presence of $H_2$, gives variable results depending on the growth temperature,[43] substrate roughness,[44, 45] substrate structure,[46] position of the growth substrate relative to the precursor source,[35, 36, 43] and precursor flux.[12, 16] To date, it is unknown how growth pressure affects CVD h-BN growth on Cu using ammonia-borane. One can grow h-BN with large (~35 $\mu m^2$) grain sizes under atmospheric pressure CVD conditions, provided that the precursor flux is kept very low, an electropolished Cu surface is used, and the substrate temperature is kept higher than 1000 °C.[45] Nevertheless, little mechanistic information is available about the CVD growth of h-BN from ammonia-borane—especially at different growth pressures—although a report suggests that ammonia-borane derived growths on Cu proceed by surface catalysis.[16]

Here, we examine conditions under which crystalline films of h-BN can be grown from ammonia-borane and $H_2$ by CVD. We find that low pressure CVD (LPCVD) gives uniform, planar h-BN films, as ascertained by scanning electron microscopy (SEM), atomic force microscopy (AFM), X-ray photoelectron spectroscopy (XPS), Fourier transform infrared (FTIR) spectroscopy, Raman spectroscopy, and scanning tunneling microscopy (STM) and spectroscopy (STS). While holding the precursor temperature and carrier gas flows constant, increasing the growth pressure significantly alters the properties of the h-BN, producing a thicker, more disordered film with non-planar, $sp^3$ components. Further, higher precursor flux conditions in LPCVD growth give thicker, nanocrystalline h-BN films, showing the importance of the ratio of precursor to $H_2$ in h-BN CVD growth. As demonstrated for graphene growth,[47-50] CVD growth of h-BN is sensitive to the Ar/$H_2$ background pressure.

## EXPERIMENTAL SECTION

**Chemical Vapor Deposition (CVD) of Hexagonal Boron Nitride (h-BN) on Cu Foil.** Growth experiments are conducted in a retrofitted Atomate CVD furnace (see Supporting Information Figure S1). Care must be taken to eliminate adventitious carbon sources from the growth chamber, otherwise h-BNC[69] or defective G/h-BN heterostructures will result; the presence of these impurity phases can be detected by Raman spectroscopy. The substrates are 0.001 inch thick (0.0254 mm) 99.8% purity Cu foils (Alfa Aesar), which are rinsed before use with a 10:1 $H_2O$:HCl solution as previously described[51-53] and annealed for 2 hr under Ar/$H_2$ (500 sccm Ar / 100 sccm $H_2$) at 1000 °C. Basic Copper (BC) of similar purity[54] was used a Cu substrate for some growths; we discriminate these substrates accordingly. The annealing step increases the Cu grain size and lowers the number of BN nucleation sites.[44, 45] The precursor $H_3N$–$BH_3$ (Aldrich) is transferred under $N_2$ into a stainless steel ampoule, minimizing water exposure of the hygroscopic $H_3N$–$BH_3$. To transport the precursor into the furnace, the ampoule is heated to ~95 °C, and volatilized material is swept into the furnace by a 4:1 $Ar/H_2$ carrier. The h-BN films are grown at 1000 °C in an $Ar/H_2$ background at the different pressures indicated. After 25 min of film growth, the samples are cooled at ~20 °C min$^{-1}$ under Ar at a flow rate of 500 sccm. See the Supporting Information for further discussion of the conditions that result in a high mass (HM) flux for the precursor temperatures used. For further characterization, the h-BN films are transferred from the Cu substrate to 90 nm $SiO_2$/Si wafers using methodologies detailed elsewhere.[53, 55]

**Sample Annealing after Transfer.** Post-transfer annealing of the samples to remove polymer residues from the poly(methyl methacrylate) or polycarbonate transfer handles also used the same furnace as h-BN growth, using a quartz tube dedicated to sample annealing. Attempts to anneal transferred h-BN films under conditions similar to those used for annealing graphene—namely, 1 hour in $Ar/H_2$ at 400 °C—leads to pitting and etching of the films, as has been reported.[22] Therefore, post-transfer sample anneals were carried out in air at 550 °C as described by others;[56] h-BN is known to be oxidation resistant under these conditions.[6, 57]

**Lithographic Patterning.** To obtain lithographically patterned h-BN samples, we defined a square array in the h-BN film by UV photolithography, using a conventional TEM grid as a mask and an $O_2$ plasma to etch. Polymethylglutarimide (PMGI; MicroChem) was spun at 3500 RPM for 30 s and cured at 165 °C for 5 min. Shipley 1813 photoresist (MicroChem) was spun on top of the cured PMGI at 5000 RPM for 30 s. The photoresist was soft baked at 110 °C for 70 s, flood exposed to UV (i-line, 365 nm) through a TEM grid for 4 s on a Karl-Suss aligner, and developed for 50 s in MF-319 (MicroChem). Using the patterned photoresist as a mask, we etched the samples in an $O_2$ plasma for 1 min under 20 sccm of $O_2$ at 100 mTorr throttle pressure and ~90 W power. After the $O_2$ etch, the samples were soaked in hot (~50 °C) Remover PG (nominally n-methylpyrrolidone) for 20 min.

**Scanning Electron Microscopy (SEM).** The h-BN films were examined immediately after growth on the Cu foil by an FEI environmental SEM operating at 5 kV. All images were taken using an ultrahigh-definition mode, which increases the dwell time and the beam current. We maintained similar values for the brightness and contrast during image collection, so that the images in Figure 1 can be compared.

**Atomic Force Microscopy (AFM).** Most AFM images were collected in tapping mode with ~300 kHz Si cantilevers on a Bruker AFM with a Dimension IV controller. Scan rates were slower than 2 Hz, and sampling was at least 512 samples per line by 512 lines; most of the scans were 1024 x 1024 images. Images with low noise and stable phase were selected for analysis. Images were de-streaked, plane fit, and analyzed using Gwyddion.[58] Root mean square (RMS) roughness values were determined using Gwyddion and by means of an algorithm written in MATLAB. Autocorrelation values were also determined and fit in Gwyddion. Some AFM images were taken on an Asylum Research MFP-3D AFM in tapping mode using ~300 kHz resonant frequency Si cantilevers (NSG30 AFM tips from NT-MDT).

**X-ray Photoelectron Spectroscopy (XPS).** A Kratos ULTRA XPS with a monochromatic Kα Al X-ray line was used to collect data. We fitted all sub-peaks with Shirley backgrounds and Gaussian-Lorentzian (GL) mixing. The amount of GL character was optimized (i.e., not fixed) in our fits, so as to lower the chi-squared value and be representative of the true chemical state of the sub-peak in question. To prevent sample charging, samples were mounted on a conducting stage using conductive tape or a metal clamp and were exposed to a flood gun during data collection. All core levels were charge corrected to the adventitious $sp^2/sp^3$ C 1s peak at 284.8 eV.

**Fourier Transform Infrared (FTIR) Spectroscopy.** FTIR spectra of the h-BN films while still on the Cu foil growth substrate were collected on a Thermo Nicolet NEXUS 670 FTIR with a Smart iTR Attenuated Total Reflectance (ATR) Sampling Accessory with a ZnSe window. All spectra were acquired in air. Before measuring an h-BN spectrum, an air background with no sample on the ZnSe window was collected and used. Each spectrum was the sum of 256 scans at least 2 wavenumber resolution.

**Raman Spectroscopy.** Raman spectra were acquired on a Horiba LabRAM HR 3D-capable imaging system at 532 nm. Data were collected with an 1800 lines/mm grating, a 100× (0.8 NA) objective, and a power level below ~10 mW. The Raman cross-section[59] of h-BN is low at 532 nm and, to improve signal-to-noise ratio and avoid the fluorescent background of the Cu foil, the Raman spectra measurements were performed on h-BN films that had been transferred to $SiO_2/Si$ substrates. Raman mapping data were acquired around the $E_{2g}$ band position (~1370 cm$^{-1}$)[60, 61] using a minimum array of 100 Raman point spectra at ~5 μm point spacing. Each spectrum consisted of an average of four to six individual measurements made at the same location, each with a 45 to 60 s acquisition time. The $E_{2g}$ mode and the higher order (~1450 cm$^{-1}$) Si 3TO mode intrinsic to the SiO$_2$/Si substrate were each fit with a single Lorentzian function using a Levenburg-Marquardt fitting algorithm in Fityk.[62]

**Scanning Tunneling Microscopy (STM).** Our experiments employed a homebuilt, room-temperature ultrahigh vacuum scanning tunneling microscope (UHV-STM)[63] with a base pressure of ~3×10$^{-11}$ Torr and electrochemically etched W and PtIr tips. Some of the tips were sharpened using field-directed sputter sharpening.[64] We scanned the samples in constant-current mode, in which the feedback electronics controlled the tip height in order to maintain a current set point, while rastering the tip across the surface. The STM tip was grounded through a current amplifier, and the tunneling bias was applied to the sample. For the constant-spacing scanning tunneling spectroscopy (STS) measurements, the tip was stopped at predetermined locations, the tip feedback was turned off, and the tip-sample bias was swept through the specified range while recording the tunneling current.

**Ultraviolet-Visible (UV-vis) Absorption Spectroscopy.** A Shimadzu UV-1650 PC instrument was used to collect UV-vis transmission spectra at an incidence angle of approximately ~60°. h-BN films were transferred onto UV-transparent quartz slides (SPI, part number 01020-AB). The transmission spectra were acquired with respect to a blank quartz slide reference.

**RESULTS**

We have grown thin films of h-BN by CVD on polycrystalline Cu foils at 1000 °C. Our growths last for 25 min in the presence of H$_2$ using the precursor ammonia-borane. Although temperatures above ~1170 °C are required to produce crystalline h-BN from ammonia-borane in the solid state,[34, 42] metal catalysts can reduce this threshold to the ~1000 °C temperatures used in our experiments.[26, 65]

The precursor reservoir is kept at ~95 °C during deposition runs. At this temperature ammonia-borane decomposes slowly to generate primarily H$_2$ and monomeric aminoborane, with negligible (i.e., undetected by mass spectrometry) amounts of borazine and diborane.[34, 42] Typically, upon opening the valve to the precursor to begin growth, the pressure in the chamber increases slightly by ~0.05 Torr; this pressure spike dissipates over about ~1 min. For some of the deposition runs, especially if the reservoir temperature is ~100 °C, the pressure increase upon opening the valve to the reservoir is somewhat higher (up to several Torr), undoubtedly due to build-up of H$_2$ and ammonia-borane decomposition products in the reservoir. We refer to these higher ammonia-borane partial pressure spikes as high mass (HM) flow conditions (see Supporting Information for further discussion).

A 4:1 mixture of Ar:H$_2$, with flow rates of either 400:100 standard cubic centimeters per minute (sccm) or 200:50 sccm, is used as a carrier gas to transport the precursor to the growth chamber. Three pressure regimes were investigated: (1) low pressure CVD conditions (LPCVD), for which the total Ar/H$_2$ background pressure, $P_{TOT}$, is 2 Torr ($P_{H2}$ = 0.4 Torr) or less; (2) medium pressure conditions, with either $P_{TOT}$ = 20 Torr ($P_{H2}$ = 4 Torr) or $P_{TOT}$ = 200 Torr ($P_{H2}$ = 40 Torr); and (3) atmospheric CVD conditions (APCVD), for which $P_{TOT}$ = 760 Torr ($P_{H2}$ = 152 Torr). We estimate the $P_{H2}/P_{H3N-BH3}$ ratio over the growth surface is ~50 for our LPCVD conditions. The $P_{H2}/P_{H3N-BH3}$ ratio increases for higher $P_{TOT}$ values and decreases for HM flow conditions.

Figure 1a shows a large-area SEM image of an h-BN film grown on Cu under LPCVD conditions. The closely-spaced (ca. 0.025 μm) striations running at approximately a −20° angle from vertical in the image indicate the step flow direction of the underlying Cu surface. These steps, which result from the mismatch in thermal expansion coefficients between h-BN and Cu, form only if the overlayer is planar and has a well-ordered crystalline structure.[66] The faint striations running diagonally from upper right to lower left are attributed to thermally induced wrinkles in the h-BN overlayer.[13, 16, 32, 36, 67] Figure 1b is a smaller-area SEM image of another region of the same LPCVD grown h-BN sample. The prominent feature that approximately bisects the image vertically is a Cu grain boundary. Here, an h-BN wrinkle crosses the Cu grain boundary; similar behavior has been seen for graphene.[28]

When growth is conducted at medium pressures of background gas, the Cu step flow features and h-BN wrinkles are absent and the surface appears morphologically rough (Figure 1c). There is also a high density of nanoparticles in the image. Similar results are obtained at a growth pressure of 200 Torr (Figure 1d); the surface is rough, and Cu step flow features are absent. Under APCVD conditions, the h-BN film is non-planar and exhibits disordered surface features in both the larger-area (Figure 1e) and smaller-area (Figure 1f) SEM images. There are no previous reports of such disordered surface features for APCVD grown h-BN. There are no obvious Cu step flow features or h-BN wrinkles, suggesting that this film is thicker than those grown under LPCVD conditions. For additional SEM images of partially grown, sub-monolayer h-BN, LPCVD, and APCVD grown h-BN, see Supporting Information Figure S3 and S4.

Figure 2 shows AFM images of the h-BN films as a function of the Ar/H$_2$ background pressure ($P_{TOT}$), along with height profiles across a film boundary generated by lithographic etching. Figure 2a shows an h-BN film grown at $P_{TOT}$ = 1.2 Torr (LPCVD regime); the film edge is indicated by the dashed, blue line. The film thickness of $0.8 \pm 0.1$ nm corresponds well to 1 to 2 h-BN layers,[9, 10] and the root-mean-square (RMS) roughness is 0.58 nm. The larger RMS roughness relative to the film thickness likely stems from entrapped water and polymeric contaminants introduced in the h-BN film transfer.[53] When a film grown under similar LPCVD conditions ($P_{TOT} = 2.0$ Torr) is annealed in Ar/H$_2$ at 400 °C, the film becomes smoother (0.45 nm RMS roughness) and etch tracks are formed (Figure 2b). The step height of the film after annealing is $1.0 \pm 0.3$ nm.

Figure 2c and 2d shows AFM topographs of two different films also grown under LPCVD conditions (2.0 Torr), except that there was a higher than normal flux from the reservoir during growth (HM conditions). The higher flux was a consequence of keeping the reservoir at ~100 °C vs. 95 °C; at the higher temperature, the precursor evolves up to 100 times more H$_2$ and volatile B- and N-containing species (see Supporting Information for further discussion). The resulting h-BN films are both thicker ($3.2 \pm 1.4$ nm) and rougher (1.51 nm RMS roughness) than those grown under LPCVD conditions when the precursor reservoir is not overheated. When grown at intermediate background pressures of Ar/H$_2$ gas ($P_{TOT} = 20$ Torr) but at normal flux from the reservoir, the films are also thick ($3.4 \pm 0.6$ nm) and rough (3.20 nm RMS roughness). These values resemble those for the sample grown at 2.0 Torr under HM conditions, underscoring the importance of controlling the precursor flux during growth.

For intermediate growth pressures ($P_{TOT} = 20$ Torr), the h-BN thin films template Cu substrate morphology, as apparent from the former Cu annealing twin shown in the topograph of Figure 2e. Vicinal Cu surfaces are known to lead to more defective, thicker graphene growth,[54] potentially explaining the enhanced h-BN growth on the twin. At higher growth pressures ($P_{TOT} = 200$ Torr), the film in Figure 2f and 2g has a step height of $10.1 \pm 0.9$ nm (1.53 nm RMS roughness), with large protrusions from transfer induced PMMA residuals.[53] Under APCVD conditions (Figure 2h), the h-BN films are rougher still (1.64 nm RMS roughness) and possess inhomogeneous depressions with contours corresponding to the morphology seen in the SEM images (Figure 1e and 1f).

The step height contours for different growth pressures shown in Figure 2i indicate that the film thickness—and thus the growth rate—increases monotonically with increasing Ar/H$_2$ background pressure, except at the very highest (APCVD) pressures. The unexpected decrease in film thickness for h-BN grown under APCVD conditions seen in Figure 2h could result from a slowed chemical reaction between the O$_2$ etching plasma and the APCVD grown h-BN thin film. Lowered O$_2$ plasma activity is consistent with a different structural (i.e., disordered) and chemical (i.e., higher N content) character for the APCVD versus LPCVD films, as detailed below. Regardless, analysis of Cu substrate photoelectron attenuation (see Supporting Information) from the h-BN overlayer allows us to estimate the APCVD film thickness at 17.8 ± 1.1 nm.

Figure 3a and 3b shows XPS photoelectron data in the B 1s and N 1s core level regions, respectively, for h-BN samples grown at $P_{TOT}$ = 2.0, 20, 200, and 760 Torr. The Supporting Information gives XPS data for a film grown at 1.2 Torr LPCVD and an additional film grown at 760 Torr (Figure S6). Table 1 summarizes the peak binding energies and FWHM values. The principal B 1s feature has a binding energy (BE) of ~190.5 eV, characteristic of bulk BN (both hexagonal and cubic phases).[68] The plots contain XPS data for h-BN films grown at 2.0 Torr on Cu foil from two different vendors, Alfa Aesar and Basic Copper (BC).[54] The data for both films possess similar line shapes and peak position BEs. For all the films, a broad π plasmon loss occurs at ~199.5 eV,[68, 69] demonstrating that the films are hexagonal and not cubic in phase.

For films grown at Ar/$H_2$ background pressures above 2.0 Torr, there is an additional B 1s feature at ~191.1 to 191.6 eV, as determined by the deconvolution of the core level spectra. This feature is characteristic of $sp^3$ B centers, such as those found in $sp^3$ rich amorphous BN films[69] and in polyaminoborane (BE = 191.1 eV).[70] The FWHM of this $sp^3$ B component increases as the background pressure increases, with the largest values measured under APCVD conditions (Table 1, Supporting Information Table S5-S7). The main N 1s XPS feature has a BE of ~398 eV, which is also characteristic of h-BN.[68] The N 1s features shown in Figure 3b do not significantly broaden or shift in BE as the growth pressure increases. However, some APCVD h-BN thin films have a higher BE sub-peak in the N 1s core level (see Supporting Information Figure S6), corresponding to $sp^3$ N (see Supporting Information Figure S6).[69] While the films consist mostly of $sp^2$ h-BN, at higher growth pressures they possess a small $sp^3$ component.

As the Ar/$H_2$ background pressure increases, the B:N stoichiometry of the films—as determined from the total areas of the B 1s and N 1s core levels—decreases from 1:1.0 ($P_{TOT}$ = 2.0 Torr) to 1:0.81 ($P_{TOT}$ = 200 Torr). At $P_{TOT}$ = 760 Torr, the B:N stoichiometry returns to a 1:1 ratio. The spectra suggest that the films consist of 1:1 h-BN that is increasingly mixed with a B-rich $sp^3$ component at background pressures up to 200 Torr. If we examine only the area under the $sp^2$ components for the B 1s and N 1s core levels, the h-BN films grown at $P_{TOT}$ ≤ 200 Torr are stoichiometric (namely, a 1:1 ratio of the areas of the B and N $sp^2$ components). Conversely, the APCVD h-BN film in Figure 3a and 3b has a $sp^2$ B:N stoichiometry of 1:1.22, implying that the APCVD grown h-BN is in a different chemical state. Thus, higher $Ar/H_2$ background pressures lead to the generation of disordered surface features (Figure 1e, 1f, and 2h) and change the chemical state in the h-BN thin films.

Depth profiles generated from time-of-flight secondary ion mass spectroscopy (TOF-SIMS) data as a function of sputtering time suggest that B—but not N—diffuses into the Cu substrate during growth of an h-BN film at a $Ar/H_2$ background pressure of 2.0 Torr (Figure 3c). We track the nitrogen in the Cu subsurface by the Cu+N mass spectra in Figure 3c, since elemental nitrogen is challenging to observe in $Cs^+$ based TOF-SIMS profiling. The Cu+N mass spectra are significantly weaker than the B mass spectra, and they track the subsurface $H_2$. These observations suggest that the N does not diffuse into the Cu substrate. The results are not surprising, especially at a growth temperature of 1000 °C. Whereas B is soluble in copper, N is not,[71, 72] although there may be some diffusion of N atoms into Cu grain boundaries, as seen for Ni and Co foils.[73] Sub-surface B during h-BN synthesis is expected, as reported for borazine-derived growth[74] of h-BN films. Thus, low pressure, ammonia-borane derived h-BN syntheses proceed by both bulk precipitation and surface catalysis. Further, these findings are consistent the recent observation of bulk precipitation for CVD h-BN growth on Ni and Co foils.[73, 75]

Figure 4a shows FTIR spectra for h-BN films grown under a variety of conditions; a spectrum of the bare Cu foil is also presented. The h-BN samples all have a peak near ~824 $cm^{-1}$ for the $A_{2u}$ out-of-plane h-BN vibration.[20, 60, 76] The ATR accessory used does not permit observation of the $E_{1u}$ in-plane vibrational mode of h-BN near ~1367 $cm^{-1}$.[60, 76] The intensity of the $A_{2u}$ peak is similar for the LPCVD h-BN samples grown 1.2 Torr and 2.0 Torr, but the $A_{2u}$ peak intensity increases—indicating thicker films—if the growth time is doubled to 50 minutes or the growth is conducted under HM flow conditions. A thicker h-BN film would lead to increased signal intensity.

The intensity of the ~823 $cm^{-1}$ peak for the h-BN film grown at 760 Torr is slightly smaller than that seen for the LPCVD samples, which is unexpected given that AFM measurements show that this sample is thicker than the LPCVD samples (Figure 2). The altered morphological and chemical structure of the APCVD grown film may be responsible for this effect. Although weak, there is a feature near ~794 $cm^{-1}$ that is characteristic of disordered h-BN films.[34] In Figure 4b, the spectrum for the APCVD grown film also shows three distinct peaks at ~1144, ~1207, and ~1271 $cm^{-1}$ on top of a broad baseline. Byproducts generated by thermolysis of the $H_3N–BH_3$ precursor possess peaks in this range.[34, 77] These peaks corroborate the disordered, polymeric surface morphology (Figure 1 and 2) and the sp$^3$ B and N chemical components (Figure 3) seen in the APCVD h-BN thin films.

Figure 5a shows Raman spectra for h-BN films grown at $P_{TOT}$ = 1.2, 2.0, 20, 200, and 760 Torr. The spectra show both the h-BN $E_{2g}$ band[60, 61] (which for bulk samples appears at ~1366 cm$^{-1}$) and the Si 3TO mode[78] at ~1450 cm$^{-1}$. The intensity of the Si 3TO mode from the substrate decreases as the growth pressure increases; this correlates with the increased h-BN film thickness (Figure 2i). As the h-BN films becomes thicker, the $E_{2g}$ mode frequency decreases from 1370.2 cm$^{-1}$ for $P_{TOT}$ = 1.2 Torr to 1368.6 cm$^{-1}$ for $P_{TOT}$ = 200 Torr. Moreover, the $E_{2g}$ frequency is essentially unchanged for thicker growths (1369.1 cm$^{-1}$ at $P_{TOT}$ = 760 Torr); see the histogram plots in Supporting Information Figure S9. The $E_{2g}$ frequencies of the thickest films are higher (i.e., less bulk-like) than expected,[61] likely reflecting either a change in the chemical structure or regions of inhomogeneous strain in the high pressure grown films.[33, 79]

The full width at half maximum (FWHM) of the $E_{2g}$ band is less sensitive to strain and is a good measure of h-BN crystallite ordering.[80] The spectra reveal that the FWHM of the $E_{2g}$ mode—as fitted to a single Lorentzian—increases with growth pressure (Figure 5b). The FWHM values of 19.7 cm$^{-1}$ ($P_{TOT}$ = 1.2 Torr) and 19.2 cm$^{-1}$ ($P_{TOT}$ = 2.0 Torr) are similar to those reported for monolayer and bilayer h-BN films exfoliated from sintered crystals.[61] Conversely, the FWHM values of 28.3 cm$^{-1}$ ($P_{TOT}$ = 20 Torr), 24.5 cm$^{-1}$ ($P_{TOT}$ = 200 Torr), and 25.1 cm$^{-1}$ ($P_{TOT}$ = 760 Torr) indicate that the h-BN films lose long range order and become defective[80] when grown at high Ar/H$_2$ background pressures.

Since the h-BN thin films are grown on Cu, scanning tunneling microscopy and spectroscopy (STM/S) is a straightforward way to measure their electronic band gaps. Figure 6a and 6b shows STM topographs of thin h-BN films on Cu grown under LPCVD conditions at 2.0 Torr. The STM image in Figure 6a is relatively streaky, and these h-BN films required high tip-sample bias conditions for stable scanning. There is no evidence of Moiré interference patterns between the h-BN film and the Cu substrate; instead, the only features evident are the Cu substrate terraces. These terrace steps in Figure 6b are likely related to an h-BN induced reconstruction of the Cu surfaces.[66] Still, the roughened Cu morphology[54] can produce h-BN grain boundaries, which in turn will have more armchair edges[81] from the obtuse angles of Figure 6b. The need for these bias conditions, the lack of finer Cu substrate detail through the h-BN, and the missing Moiré superstructures indicate that this film is thin, but certainly more than one layer thick (i.e., 2 to 3 layers). Furthermore, these findings are consistent with the AFM step height data (Figure 2i, Supporting Information Figure S5) and also suggest that the film possesses a wide band gap.

By contrast, the STM topograph of h-BN grown at 1.2 Torr (Figure 6c) shows a linear pattern from the Cu substrate. This linear pattern has a period of ~2 nm, similar to the pattern observed for graphene on Cu(111) after oxygen intercalation.[82] Unlike the sample in Figure 6a and 6b, scanning is stable at tip-sample biases well within the h-BN band gap. Therefore, this sample is most likely an h-BN monolayer on the Cu foil substrate. The plot in Figure 6d displays the tunneling current ($I$) versus the tip-sample bias ($V$) for individual STS spectra recorded at different locations on the h-BN film grown at $P_{TOT} = 2.0$ Torr shown in Figure 6a; the solid black line indicates the average spectrum. The average band gap of ~5.7 eV is consistent with that of bulk h-BN.[9, 21]

In order to verify that the band gap measured on the nanometer scale is also characteristic of the entire film, we also characterize the films by ultraviolet-visible (UV-vis) absorption spectroscopy. The plot displayed in Figure 6e shows UV-vis absorption spectra for h-BN films grown at $P_{TOT} = 1.0$ and 1.2 Torr after transfer to UV-transparent quartz. Analysis of the UV-vis absorption spectra by the Tauc method[12, 15, 20, 21, 83] shows that the LPCVD h-BN films have optical band gaps between 5.3 eV and 5.5 eV, which is within the expected range for h-BN[9, 21] and close to electronic band gaps measured by STS. A two point probe transport measurement of an h-BN film grown under LPCVD conditions further confirms that the films are not conducting (see Supporting Information Figure S11). Our combined STS and UV-vis absorption spectroscopic measurements ultimately support the conclusion that the LPCVD h-BN films have the electronic signature of crystalline h-BN.

## DISCUSSION

For the CVD of h-BN from ammonia-borane on Cu at 1000 °C in the presence of a 4:1 Ar/H$_2$ background gas, we have found the following: at Ar/H$_2$ background pressures of 20 Torr ($P_{H2} = 4$ Torr) or above, the growth rate increases with pressure, but the films contain larger and larger amounts of a sp$^3$ component that is similar to amorphous BN or polyaminoborane (PAB). The faster growth rate is consistent with previous studies of the CVD of graphene, showing that H$_2$ serves as a co-catalyst[49] and that the growth depends on the H$_2$ to precursor ratio.[49, 50] However, the results are in contrast to previous findings that H$_2$ etches spurious h-BN nucleation[22] and removes polymeric PAB and polyiminoborane (PIB) species,[84] as we observe h-BN, PAB, and PIB compounds in h-BN films grown at high Ar/H$_2$ background pressure. We can reconcile the contrasting observations by proposing that, at the higher $H_2$ pressures, the faster h-BN growth rates cause the Cu foil to become quickly covered, so that its catalytic activity is suppressed. As a result, the precursor does not completely decompose on top of the already deposited h-BN layers. In the absence of a catalyst, growth of h-BN from $H_3N–BH_3$ requires very high temperatures (~1500 °C).[34] Incomplete decomposition of the $H_3N–BH_3$ precursor[34, 77] explains several of the properties of the h-BN samples grown at a Ar/$H_2$ background pressure of 760 Torr: the amorphous, disordered surface morphology, the $sp^3$ components observed in the XPS data, and the extra, polymeric peaks in the FTIR spectra.

Previously, Bhaviripudi *et al.* showed that the growth of CVD graphene on Cu is demarcated into three growth regimes: surface reaction (catalysis), mixed growth, and mass transport.[47] In these growth regimes, the total pressure $P_{TOT}$ determined whether graphene growth proceeded by catalysis or otherwise. For h-BN growths on Cu at $P_{TOT} > 20$ Torr (medium pressure CVD to APCVD), we note an increase in the surface growth rate $K_s$ and a significant decrease in the mass transfer coefficient $h_g$. The decrease in mass transfer coefficient $h_g$ occurs because the coefficient is inversely proportional to $P_{TOT}$: $h_G \propto P_{TOT}^{-1} \delta^{-1}$, where $\delta$ is the boundary layer thickness, weakly dependent on $P_{TOT}$.[47] Under these conditions, $h_g << K_s$, making the mass transport flux dominant over surface reactions, ultimately eliminating the high temperature catalytic decomposition of ammonia borane. LPCVD h-BN growth conditions increase $h_g$, promoting surface catalysis ($h_g >> K_s$), growing $sp^2$ h-BN, and removing spurious PAB and PIB. Remnant borane derivatives and substrate N in-diffusion can drive local B and N gradients, influencing the shape of growing h-BN nuclei into atypical morphologies, as a recent theoretical report predicted.[85]

Finally, it is important to control the temperature of the precursor reservoir, because the rate of decomposition of $H_3N–BH_3$ to $H_2$, monomeric aminoborane, borazine, and other (mostly non-volatile) products is quite temperature sensitive.[34, 42, 77, 86] An overheated ammonia-borane precursor can lead to polymerization of the precursor, generate adventitious $H_2$, and increase the partial pressure of the precursor relative to the Ar/$H_2$ background pressure. In this HM flow growth scenario, the increased flux of $H_2$, monomeric aminoborane, and other volatile compounds will drive a thicker, nanocrystalline h-BN morphology, circumventing the catalytic activity of the Cu substrate at high temperature.

## CONCLUSION

The growth of large-arean h-BN on Cu by CVD depends critically on the background Ar/$H_2$ pressure ($P_{TOT}$). Uniform, planar thin h-BN grows under LPCVD conditions ($P_{TOT} = 2$ Torr), whereas medium pressure ($P_{TOT}$ = 200 Torr) and APCVD growth conditions ($P_{TOT}$ = 760 Torr) afford thicker films with a mixture of h-BN and partially decomposed, sp$^3$ like $H_3N–BH_3$ species. Under medium pressure and APCVD conditions, the faster growth rate leads to rougher, less crystalline films. Under APCVD conditions, amorphous, polymeric features form, which we attribute to incomplete thermolysis of the $H_3N–BH_3$ precursor due to passivation of the Cu catalyst. Moreover, growth under HM conditions shows that the growth rate depends critically on the $H_2$ to $H_3N–BH_3$ flux ratio, even under LPCVD growth conditions. Our LPCVD growths produce h-BN thin films comparable in electronic and optical quality to those recently observed for h-BN grown on sapphire.[87] By contrast, our APCVD growths are similar in variability to those reported in the literature.[21,44] To grow high-performance h-BN nanomaterials demands a clear understanding of h-BN CVD growth mechanisms. As a result, h-BN growth protocols will better control layer number, chemical heterogeneity, and crystallinity, enabling the fabrication of large area, electronic and encapsulatory h-BN heterostructures.

**AUTHOR CONTRIBUTIONS**

[†]J. C. Koepke and J. D. Wood contributed equally to this work.

**CONFLICTS OF INTEREST**

The authors declare no competing financial interests.

**ACKNOWLEDGMENTS**

This work has been sponsored by the U.S. Office of Naval Research (ONR) under grant N00014-13-1-0300, the Air Force Office of Scientific Research (AFOSR) under grant FA9550-14-1-0251 (E.P.), and the National Science Foundation (NSF) under grants CHE 10-38015 (J.W.L.), 13-07002 (L.N. and M.G.), 13-62931 (G.S.G.), and ECCS-1430530 (E.P.). J.D.W. gratefully acknowledges funding from the National Defense Science and Engineering Graduate Fellowship (NDSEG) through the Army Research Office (ARO), the Beckman Foundation, and the Naval Research Enterprise Intern Program (NREIP). This research was performed while S.W.S. held a National Research Council Research Associateship Award at the Naval Research Laboratory. We kindly thank G. Doidge, K. Chatterjee, and T. Kilpatrick for assistance in h-BN transfer. Time-of-flight secondary ion mass spectroscopy (TOF-SIMS) and X-ray photoelectron spectroscopy measurements were carried out in the Frederick Seitz Materials Research Laboratory Central Facilities at the University of Illinois. Scanning electron microscopy, Raman spectroscopy, and transmission electron microscopy measurements were performed in the Microscopy

Suite, which is part of the Imaging Technology Group at the Beckman Institute of the University of Illinois. We are indebted to S. Robinson for assistance with TEM imaging and T. Spila for help in TOF-SIMS data collection. We also acknowledge J. Kaitz for assistance in using the FTIR system.

## SUPPORTING INFORMATION

The supporting information contains the following: (1) additional discussion of the high mass flow (HM) growth conditions; (2) an image of the furnace setup; (3) characterization of the h-BN films including AFM contours of the film step heights; (4) a table summarizing the RMS roughness data for the AFM images; (5) further XPS data and details of the peak fits for the different growth pressures; (6) Raman histograms of the $E_{2g}$ peak position for different growth pressures; and, (7) a comparison of the Raman data from LPCVD grown h-BN films with normal precursor flux and under HM growth conditions. This document is available online free of charge at http://pubs.acs.org.

**FIGURES**

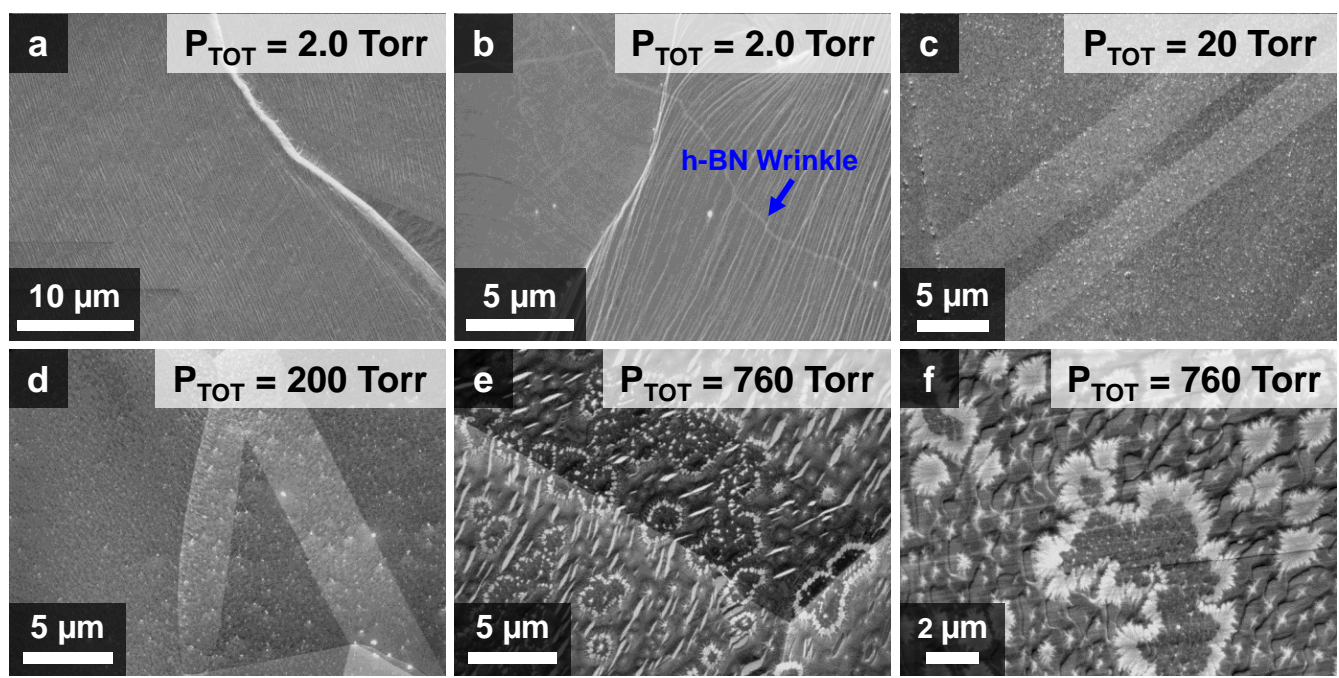

**Figure 1.** Scanning electron microscopy (SEM) imaging of h-BN on Cu at different pressures. Low-pressure h-BN growth **(a)** at a large-scale and **(b)** at a small-scale, revealing a planar h-BN film with protrusions from Cu step flow[66] and h-BN wrinkles. **(c)** Higher pressure growth, showing a loss of the Cu hillock morphology and an increase in charging. **(d)** Medium-pressure growth, with similar morphology as (c). Atmospheric pressure h-BN growth at a large-scale **(e)** and a small-scale **(f)**, with polymeric features evident. These features suggest a breakdown in Cu-mediated catalysis.

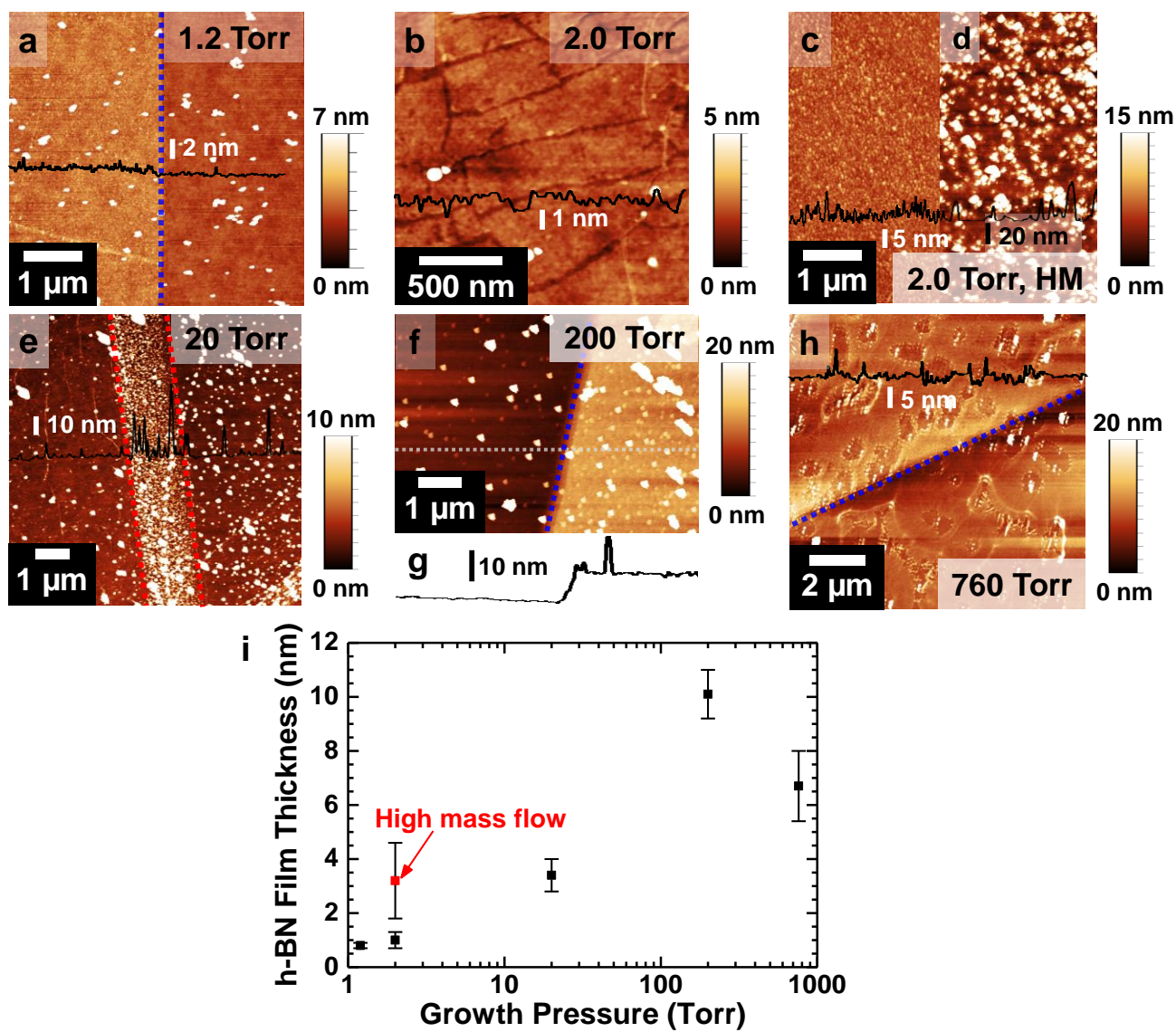

**Figure 2.** Changes in h-BN film morphology due to precursor mass flow and growth pressure. **(a)** Atomic force microscopy (AFM) images of transferred and lithographically patterned (blue line) h-BN on SiO$_2$ growth at low pressure (1.2 Torr). **(b)** Low pressure (2.0 Torr) h-BN growth with improved precursor mass flow control showing smoother morphology. **(c, d)** AFM images of transferred h-BN on SiO$_2$ grown at low pressure (2.0 Torr) but with a high mass (HM) flow of H$_3$N–BH$_3$. The HM condition gives a more nanocrystalline h-BN film, as seen in the overlaid height profile. **(e)** Medium pressure (20 Torr) h-BN growth with heightened H$_3$N–BH$_3$ catalysis on a former Cu annealing twin. AFM image **(f)** and height profile **(g)** for lithographically patterned (blue line) h-BN grown at medium pressure (200 Torr). Patterned large-area **(h)** AFM image for h-BN grown at atmospheric pressure (760 Torr). The APCVD grown film is highly inhomogeneous and rough, with polymeric depressions corresponding to the features seen in SEM. **(i)** Height profiles for the growths, showing thin h-BN films at LPCVD.

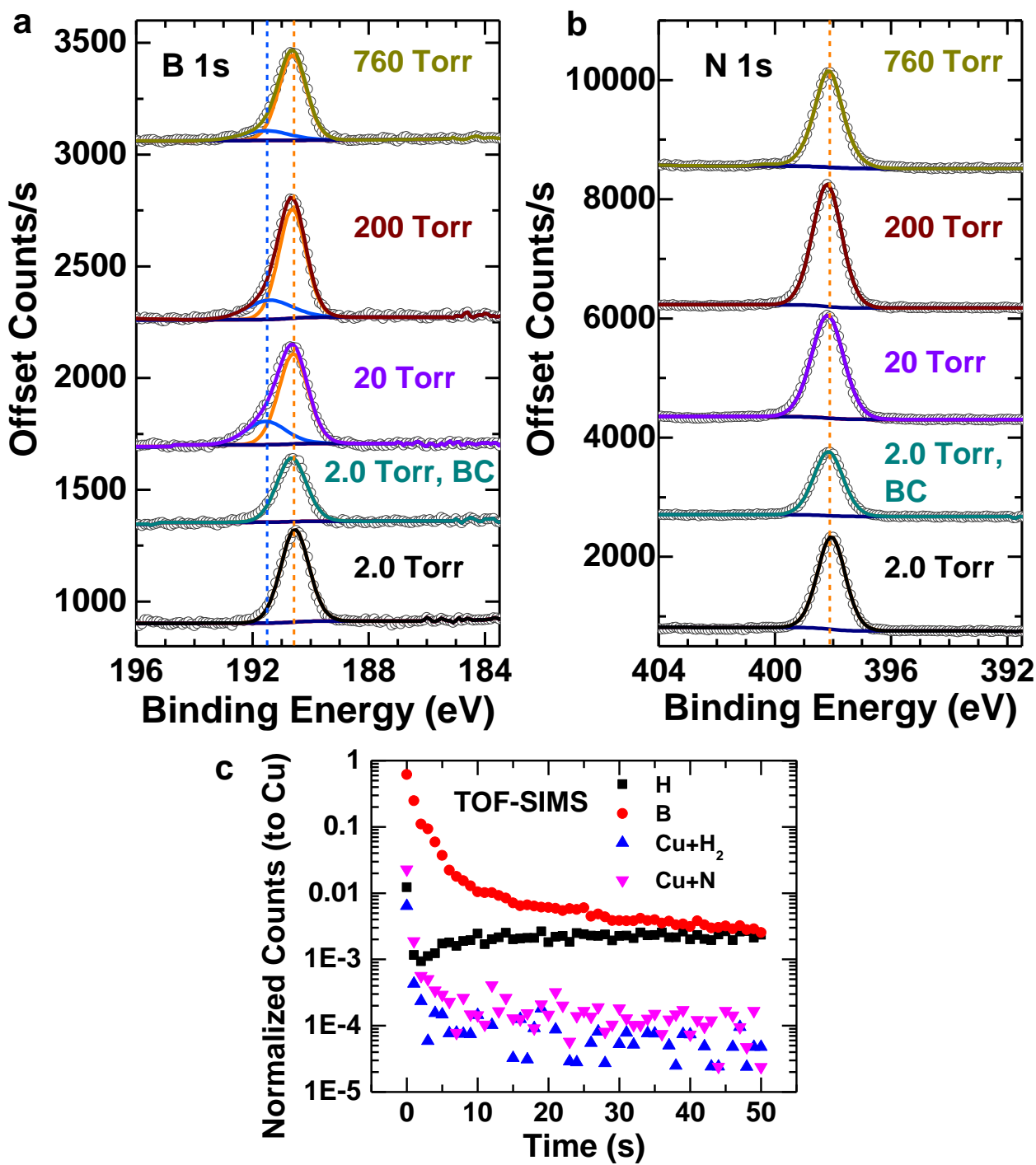

**Figure 3.** Chemical information for different h-BN growth pressure setpoints. X-ray photoelectron spectroscopy (XPS) core level data with respect to growth pressure for the B 1s **(a)** and N 1s **(b)** photoelectron (PE) lines. All growths give hexagonally structured BN, as confirmed by the main B $sp^2$ peak (orange) in the B 1s PE line. The BC label indicates Cu growth foil from Basic Copper, a different Cu source. For B 1s, a secondary $sp^3$ B peak (blue) appears and widens with increasing growth pressure, indicative of polymeric components. **(c)** Time-of-flight secondary ion mass spectroscopy (TOF-SIMS) depth profiling for low-pressure h-BN (2.0 Torr), demonstrating sub-surface B diffusion in the Cu.

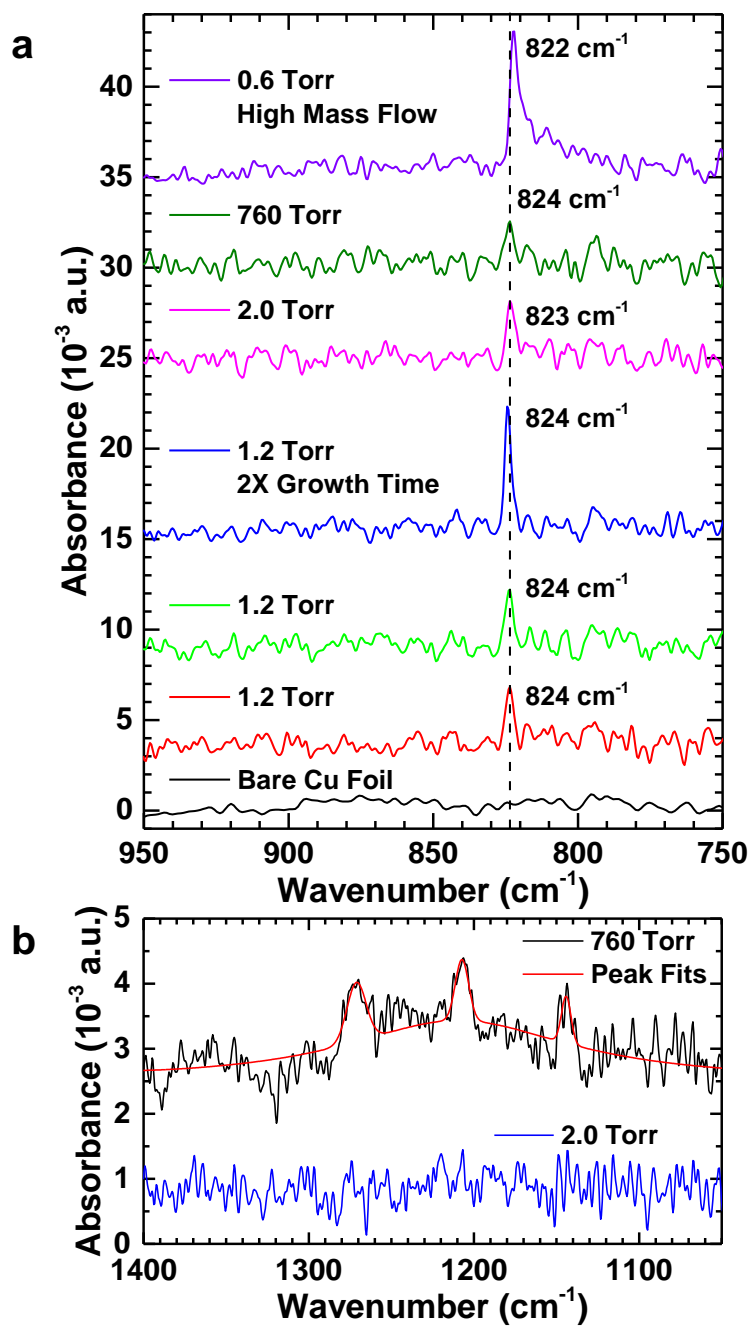

**Figure 4.** Fourier transform infrared spectroscopy (FTIR) of h-BN grown at different pressures. **(a)** Spectra for h-BN grown at different pressure in the region near the LO $A_{2u}$ mode of h-BN.[60] Spectra offset for clarity. All of the h-BN growths show a peak near ~824 cm$^{-1}$. The low intensity and peak at ~794 cm$^{-1}$ suggest a disordered film for APCVD h-BN. **(b)** High wavenumber spectra for APCVD and LPCVD (2 Torr) h-BN films. The APCVD growth shows a higher baseline *versus* the LPCVD case. Several small peaks appear about ~1200 cm$^{-1}$, demonstrating incomplete $H_3N–BH_3$ breakdown products.[34]

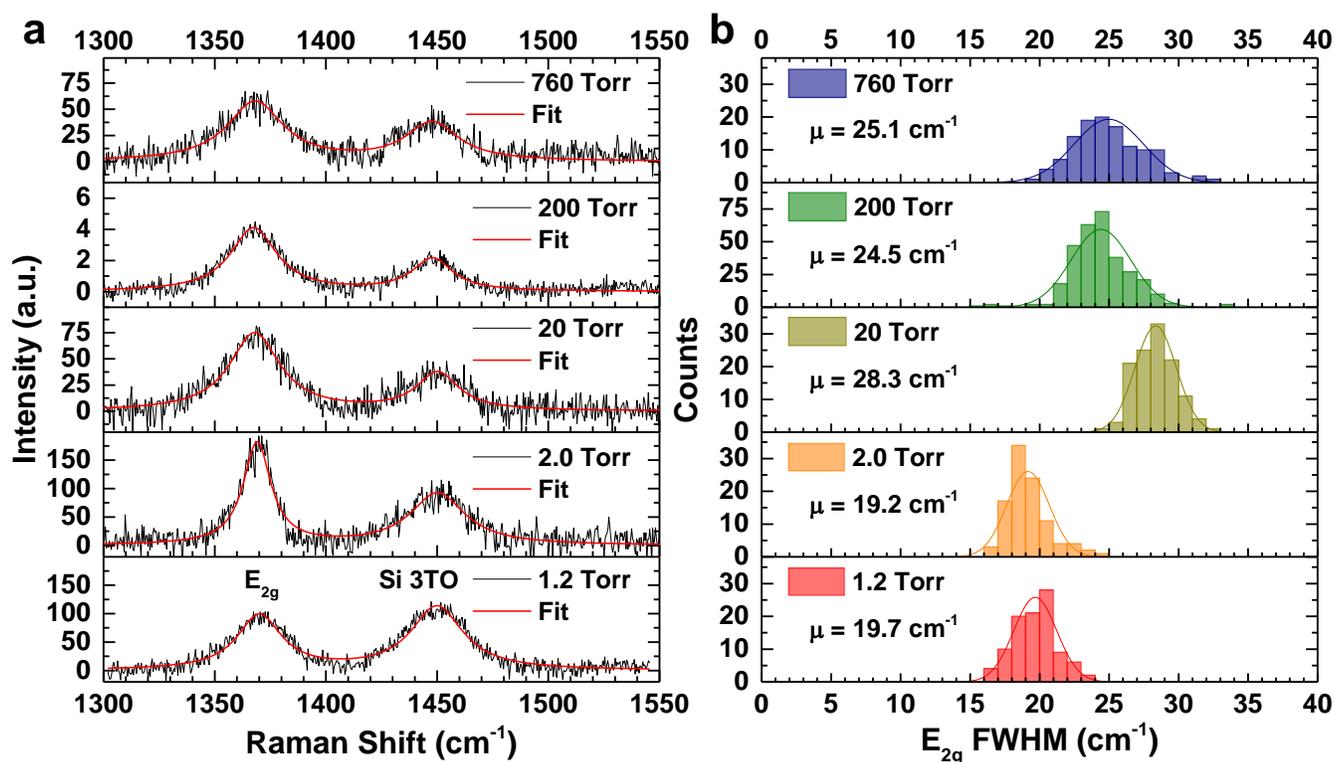

**Figure 5.** Raman spectroscopy of transferred h-BN grown at different pressures. **(a)** Representative point Raman spectra from h-BN growths, fitted by Lorentzians. The $E_{2g}$ mode of h-BN at ~1370 cm⁻¹ and the Si 3TO mode at ~1450 cm⁻¹ are apparent. **(b)** Histograms for the $E_{2g}$ band full-width at half-maximum (FWHM) using Raman mapping data from the h-BN growths. Higher growth pressure films show a larger FWHM, indicative of disorder.[80]

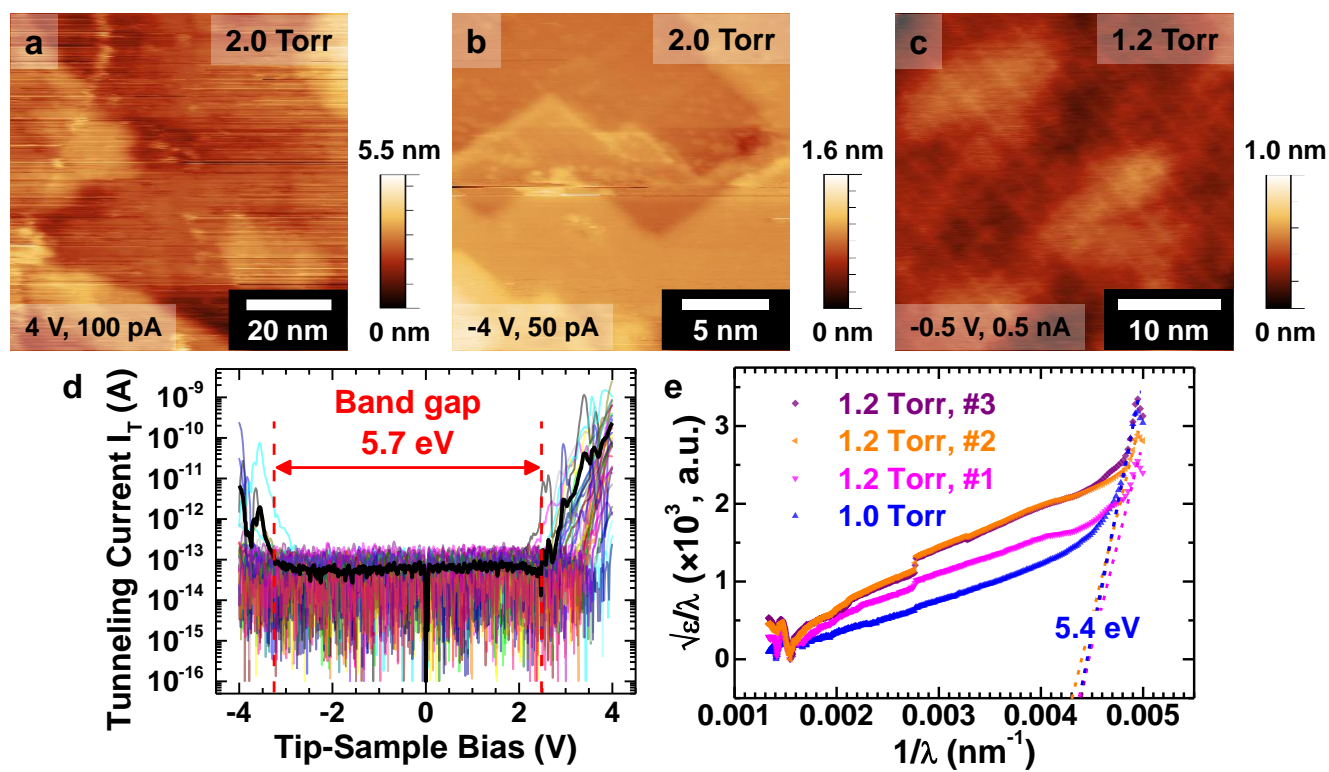

**Figure 6.** Electrical and optical data for h-BN films. **(a, b)** Scanning tunneling microscopy (STM) topographs for thin h-BN grown under LPCVD (2 Torr) conditions. The wide band gap of the grown h-BN necessitates using large tip-sample biases (±4 V) to scan the surface. Cu step edges, namely, the angular features, are also evident. **(c)** A STM topograph of lower pressure h-BN growth (~1.2 Torr). This monolayer h-BN film is thin enough to scan at smaller tip-sample biases. (d) Scanning tunneling spectroscopy (STS) data on the same low-pressure sample in **(a, b)**, showing a wide electronic band gap characteristic of h-BN. **(e)** Optical band gap extraction from low pressure (P = 1.0 and 1.2 Torr) CVD h-BN films showing band gaps near 5.4 eV.

| Growth Pressure (Torr) | sp² B | | sp³ B | | sp² N | | Stoichiometry |
|---|---|---|---|---|---|---|---|
| | BE (eV) | FWHM $\Gamma$ (eV) | BE (eV) | FWHM $\Gamma$ (eV) | BE (eV) | FWHM $\Gamma$ (eV) | y, $BN_y$ |
| 1.2 | 190.7 | 1.27 | absent | absent | 398.3 | 1.24 | 1.02 |
| 2.0 | 190.5 | 1.14 | absent | absent | 398.1 | 1.15 | 0.95 |
| 2.0, BC | 190.7 | 1.22 | absent | absent | 398.1 | 1.20 | 1.04 |
| 20 | 190.6 | 1.15 | 191.3 | 1.86 | 398.2 | 1.31 | 0.89 |
| 200 | 190.6 | 1.00 | 191.1 | 1.40 | 398.2 | 1.19 | 0.81 |
| 760 | 190.6 | 1.16 | 191.5 | 2.01 | 398.2 | 1.20 | 1.03 |
| 760 | 190.6 | 1.11 | 191.6 | 2.00 | 398.1 | 1.17 | 0.88 |

**Table 1. Summary of XPS statistics for different h-BN growths.** The binding energy (BE) and full width at half maxima (FWHM) for peak fits to XPS data are from Figure 2. Here BC is the label for h-BN growth on Cu foil from Basic Copper,[54] rather than from Alfa Aesar.

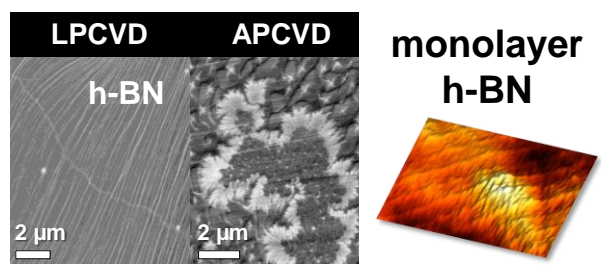

**TOC Figure**

*Supporting Information*

# Role of Pressure in the Growth of Hexagonal Boron Nitride Thin Films from Ammonia-Borane

Justin C. Koepke[1,2,3,a)†], Joshua D. Wood[1,2,3,b)†], Yaofeng Chen[1,2,3], Scott W. Schmucker[4], Ximeng Liu[1,2,3], Noel Chang[5], Lea Nienhaus[2,3,5,c)], Jae Won Do[1,2,3], Enrique A. Carrion[1,3], Jayan Hewaparakrama[1,3], Aniruddh Rangarajan[1,2,3], Isha Datye[1,2,3], Rushabh Mehta[1,2,3], Richard T. Haasch[6], Martin Gruebele[2,5,7], Gregory S. Girolami[2,5], Eric Pop[1,8], and Joseph W. Lyding[1,2,3*]

*[1]Dept. of Electrical & Computer Eng., Univ. of Illinois at Urbana-Champaign, Urbana, IL 61801*
*[2]Beckman Institute, Univ. of Illinois at Urbana-Champaign, Urbana, IL 61801*
*[3]Micro and Nanotechnology Lab, Univ. of Illinois at Urbana-Champaign, Urbana, IL 61801*
*[4]U.S. Naval Research Laboratory, Washington, D.C. 20375*
*[5]Dept. of Chemistry, Univ. of Illinois at Urbana-Champaign, Urbana, IL 61801*
*[6]Materials Research Laboratory, Univ. of Illinois at Urbana-Champaign, Urbana, IL 61801*
*[7]Dept. of Physics, Univ. of Illinois at Urbana-Champaign, Urbana, IL 61801*
*[8]Electrical Engineering, Stanford University, Stanford, CA 94305*

*a) Present address: Sandia National Laboratories, P.O. Box 5800, Albuquerque, NM 87185*
*b) Present address: Dept. of Materials Science and Eng., Northwestern Univ., Evanston, IL 60208*
*c) Present address: Dept. of Chemistry, Massachusetts Inst. of Technol., Cambridge, MA 02139*

*† These authors contributed equally.*

**Contents:**

[*] Correspondence should be addressed to lyding@illinois.edu, jkoepkeuiuc@gmail.com, and joshua.wood@northwestern.edu.

## Section S1. High ammonia borane mass (HM) flow growth conditions

The temperature of the ampoule containing the ammonia-borane ($H_3N$–$BH_3$) powder controls the precursor flux during the growth. If the precursor temperature is not controlled well, then the precursor flux during growth will vary widely. During our growth experiments, we observed that there is typically a small increase (~10 to ~100 mTorr) in the measured pressure in the growth chamber upon opening the valve between the precursor ampoule and the Ar/$H_2$ flow at the inlet of the furnace. During the first thermal cycle of the precursor, the increase in pressure upon opening the valve to the precursor can be significantly higher, from an increase in pressure by ~1 Torr to a maximum observed increase of ~10 Torr. Such a large flux of precursor byproducts flooding the chamber dramatically changes the balance between the $H_3N$–$BH_3$ byproducts and $H_2$. Previous studies of CVD growth of graphene on Cu substrates[1,2] showed that the ratio of carbon precursor to $H_2$ plays a very important role in nucleation density and grain size.

As shown in Figure 2 of the main manuscript, the resulting film thickness and roughness both increase during a growth where HM precursor conditions existed upon opening the precursor valve as compared to a growth with normal precursor flux. For the h-BN sample shown in Figure 2b, the precursor temperature was not well-controlled and varied between ~90 °C and 100 °C. For the h-BN sample shown in Figure 2c, the precursor temperature was between 99 °C and 102 °C. While a higher precursor temperature can increase the precursor byproduct flux, this HM growth condition can also occur with the precursor temperature set to the 95 °C value used for the growth of the other of the samples in this study. Typically, this HM pressure spike occurs during the first growth after re-loading the ampoule with $H_3N$–$BH_3$. The HM growth condition can be minimized during the first growth by thermally cycling the precursor to the target temperature (95 °C in this case) while pumping under vacuum for ~ 15–25 min prior to use for h-BN CVD synthesis. Subsequent h-BN growths using the thermally-cycled precursor in the ampoule will have more typical increases in the chamber pressure (usually less than 100 mTorr) upon opening the precursor valve to begin the h-BN growth step.

Raman data in Figures S10a and S10c show a comparison of Raman statistics from a normal h-BN growth at 1.0 Torr and one with HM precursor flux with a background growth pressure of 0.6 Torr. During the HM growth, the pressure spike reached 10 Torr from the 0.6 Torr Ar/$H_2$ background, and the resulting h-BN had a film thickness of ~46 nm and film roughness of 9.6 nm. While the peak position is nearly the same between the normal h-BN growth at 1.0 Torr (Figure S10b) and the HM growth at 0.6 Torr (Figure S10a), the average FWHM of the h-BN film grown under HM conditions (Figure S10c) is much higher than for the normal growth at 1.0 Torr (Figure S10d). Hence, HM growth conditions lead to h-BN films with larger $E_{2g}$ mode FWHM than for films grown under normal precursor flux conditions at the same or similar pressures. Larger FWHM for the h-BN $E_{2g}$ Raman mode indicates that HM conditions lead to h-BN films with more nanocrystalline and defective structure than their normal precursor flux counterparts.[3]

The Raman spectra for the h-BN film grown at 20 Torr (Figure 5b in the main manuscript) has an average value of ~28 cm$^{-1}$, very close to the ~29 cm$^{-1}$ average FWHM of the Raman data for the HM growth in Figure S10c. This suggests that, despite the thermal pre-treatment of the precursor immediately prior to use for this growth, the precursor flux for the h-BN sample grown at 20 Torr was also higher than expected for the precursor temperature. Given the detrimental effects of HM growth conditions, good control of the precursor temperature and preparation of the precursor are critically important for h-BN growth. While the HM growth from Figure S10a and S10c (film thickness of ~46 nm) shows that growth rate increases as the ratio between $H_2$ and the $H_3N-BH_3$ byproducts ($P_{H2}$:$P_{H3N-BH3}$) goes down, the surface never has the dendritic, disordered features observed for h-BN grown at APCVD conditions. This points to a different growth mechanism for low pressure, HM growth conditions, likely Volmer-Weber island growth. The h-BN films grown under HM conditions underscore the importance of the ratio between $H_2$ and the $H_3N-BH_3$ byproducts in determining the properties of the resulting h-BN film.

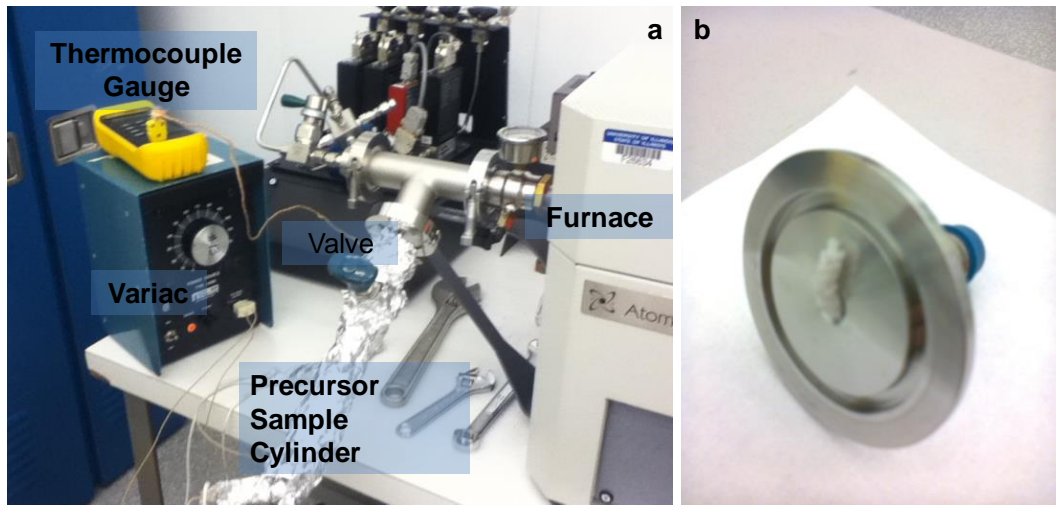

**Figure S1. Experimental setup for CVD h-BN growth. (a)** $H_3N–BH_3$ precursor cylinder and variac used to modulate the ammonia borane sublimation temperature. **(b)** Polymerized $H_3N–BH_3$ breakdown products on a stainless steel flange after CVD h-BN growth. For this run, the growth pressure was at 2.0 Torr, the growth time was 25 min, and the precursor temperature was 95 °C. Polymerization occurs in the presence of high $H_3N–BH_3$ mass flow (high sublimation). Therefore, care must be taken to control the sublimation rate carefully.

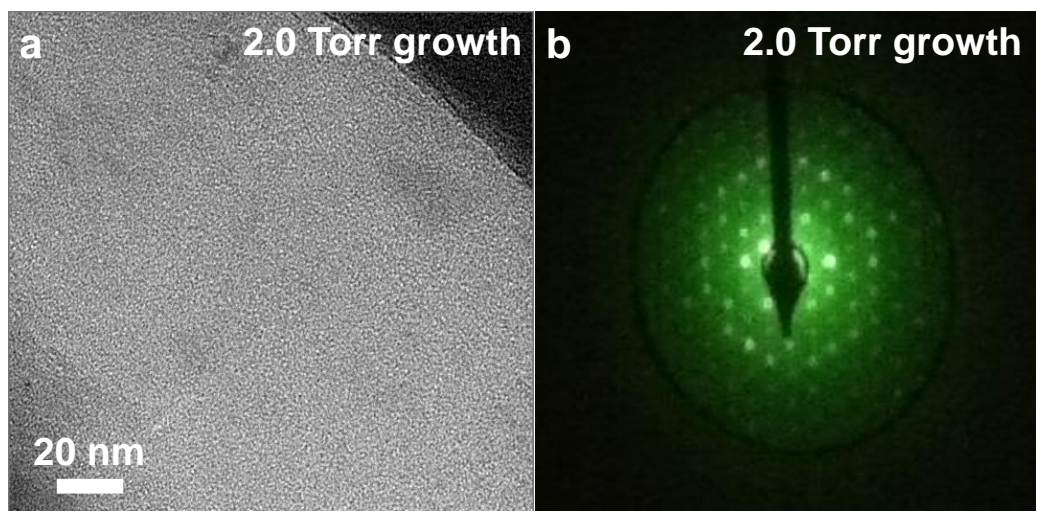

**Figure S2. Transmission electron microscopy (TEM) imaging of h-BN grown at low pressure. (a)** Bright-field TEM image of PMMA-transferred h-BN grown at 2.0 Torr. **(b)** Photograph of selected-area electron diffraction (SAED) of the area in (a), showing hexagonal, single-domain CVD h-BN. The single set of diffraction spots suggests that we have AA' stacked h-BN and not turbostratic BN (t-BN), as t-BN possesses multiple h-BN layers rotationally misoriented out of the basal plane.

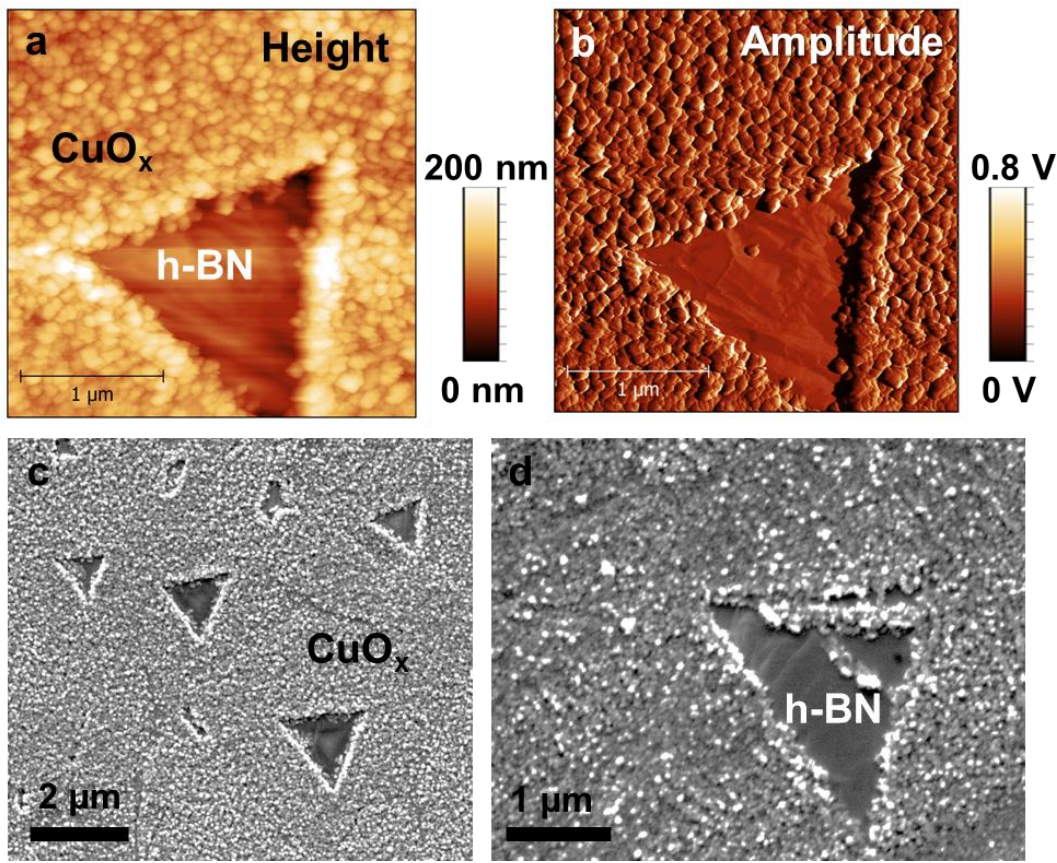

**Figure S3. Partially grown h-BN on Cu after Cu oxidation.** AFM height (**a**) and amplitude (**b**) images of h-BN on oxidized Cu. The h-BN exhibits its known triangular shape[4, 5] and is resistant to oxidation.[6,7] This Cu foil was oxidized at ~300 °C for several hours. (**c**) Large-area SEM image of the same sample in (a-b), again showing the triangular h-BN regions that protect the Cu from oxidation. (**d**) SEM zoom-in on one of the triangular h-BN features.

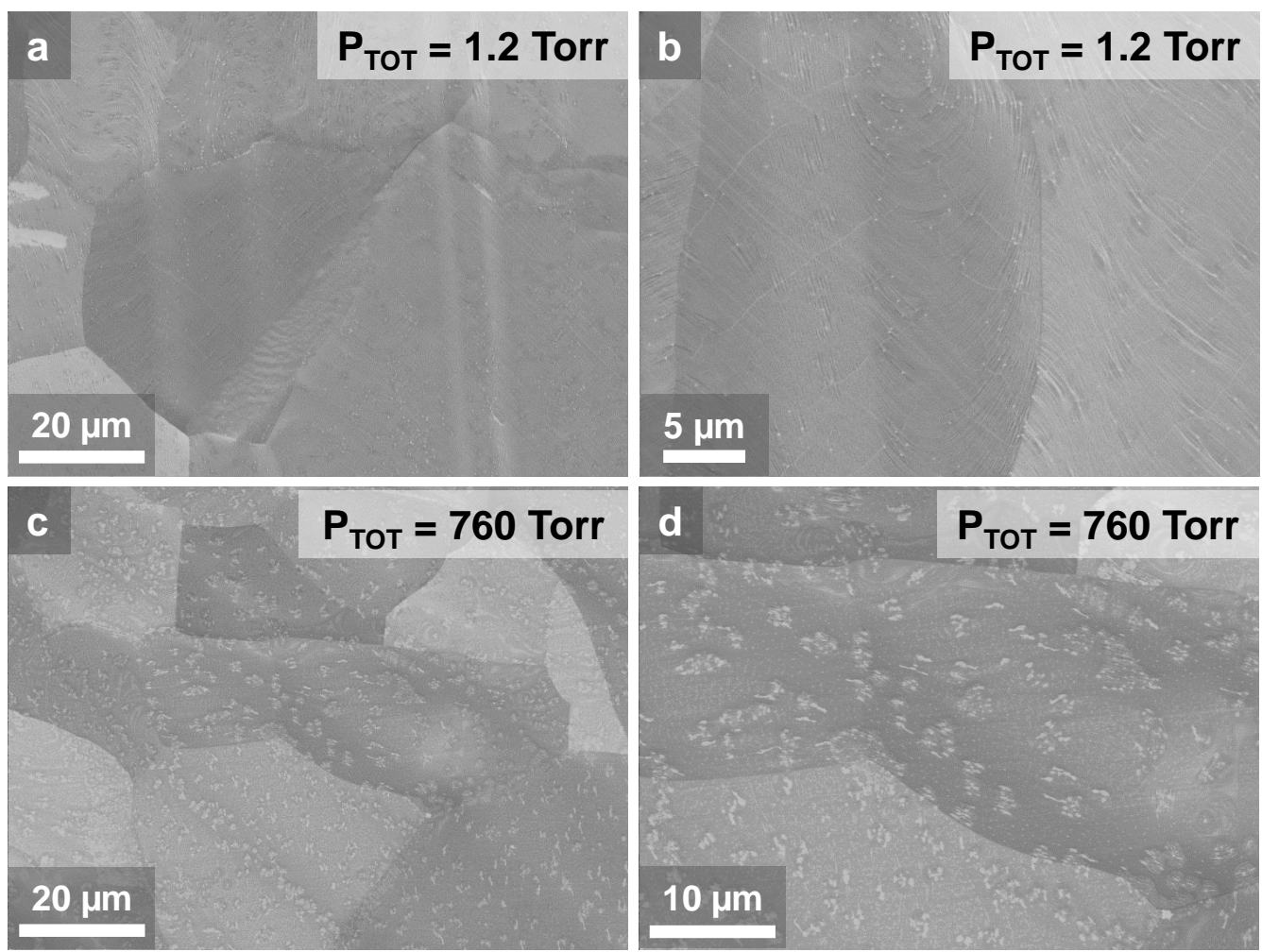

**Figure S4. Scanning electron microscopy (SEM) images of other LPCVD and APCVD h-BN growths. (a, b)** SEM images of h-BN grown at 1.2 Torr (LPCVD) shows planar, conformal growth on Cu foil similar to that for h-BN grown at 2.0 Torr (Figures 1a and 1b). **(c, d)** SEM images from a different growth at APCVD shows non-uniform, non-crystalline features on the surface, similar to that from the APCVD growth shown in Figures 1e and 1f.

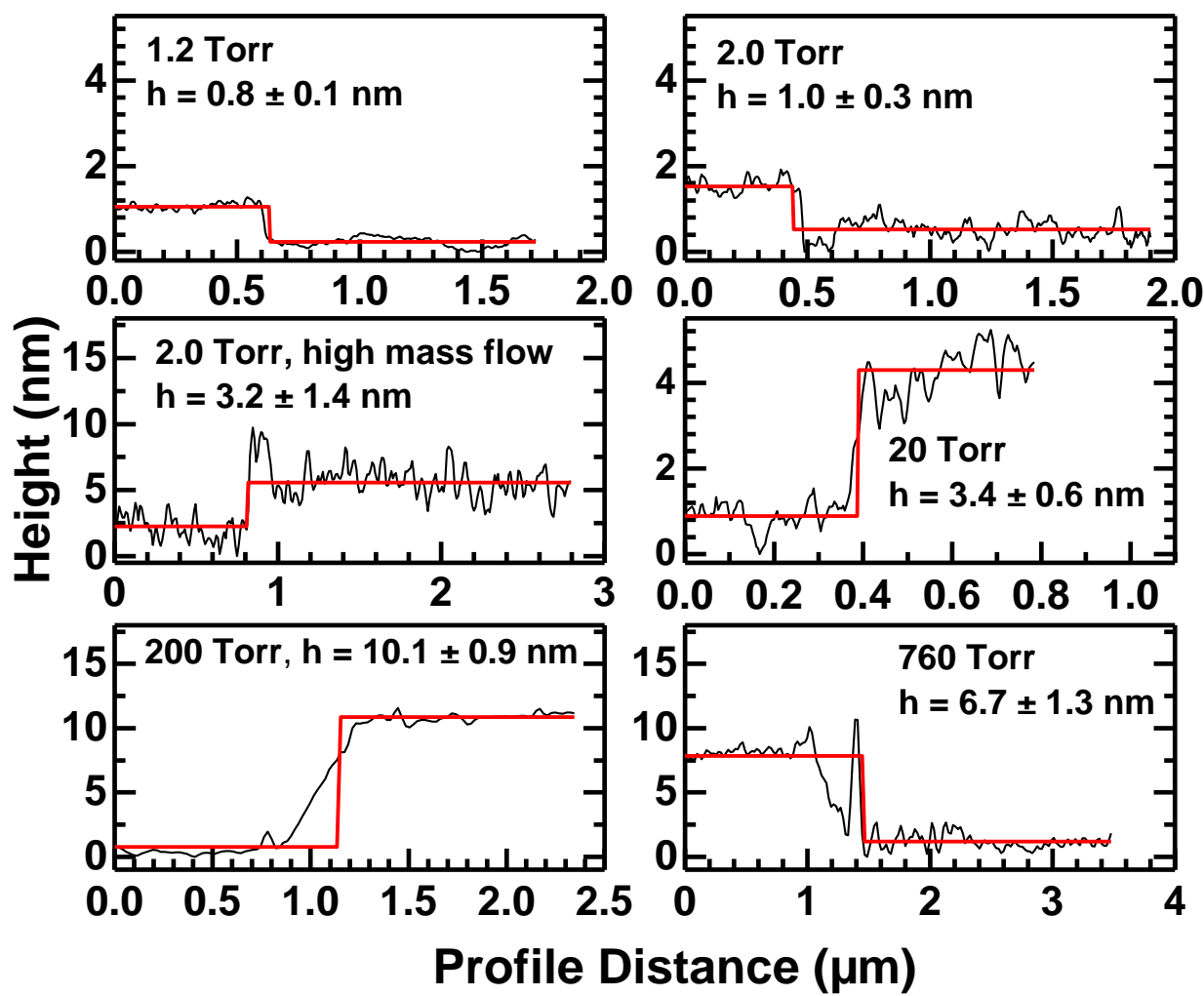

**Figure S5. AFM height profiles for different h-BN growths.** The step height contours indicate that the film thickness, and thus growth rate, increases with growth pressure. The h-BN film grown at 760 Torr did not completely etch, obscuring the actual film thickness.

| Growth ID | Growth Pressure (Torr) | RMS Roughness (nm) |
|-----------|------------------------|---------------------|
| h-BN #46 | 0.6 (HM) | 9.60 |
| h-BN #30 | 1.2 | 0.58 |
| h-BN #3 | 2.0 (HM) | 1.51 |
| h-BN #10 | 2.0 | 0.45 |
| h-BN #17 | 20 | 3.20 |
| h-BN #13 | 200 | 1.53 |
| h-BN #12 | 760 | 1.64 |

**Table S1. RMS roughness of h-BN versus growth pressure after transfer to SiO₂/Si.** AFM scans of h-BN after transfer show that the roughness of the h-BN increases with growth pressure. While the roughness of the lowest pressure growth (1.2 Torr) is higher than the 2.0 Torr growth, this is most likely due to the conformal nature of the 1 to 2 h-BN layer film on the substrate compared to the thicker film grown at 2.0 Torr. The standard error for each measurement is shown.

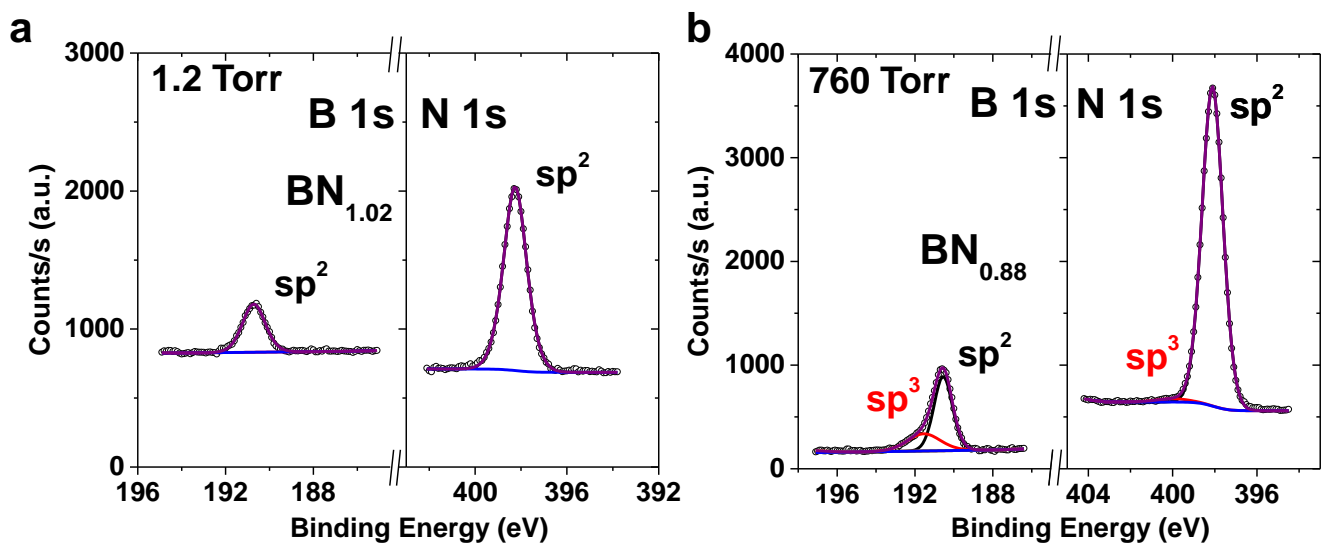

**Figure S6. X-ray photoelectron spectroscopy (XPS) core level comparison of other LPCVD and APCVD h-BN growths.** (a) B 1s and N 1s photoelectron (PE) lines for h-BN grown on Cu foil at 1.2 Torr (LPCVD conditions). (b) B 1s and N 1s PE lines for h-BN growth on Cu foil at 760 Torr (APCVD conditions). The LPCVD-grown h-BN film in (a) has no $sp^3$ component, while the APCVD-grown film in (b) shows a significant $sp^3$ component in the B 1s PE spectrum. The stoichiometry of the LPCVD-grown h-BN in (a) is nearly 1:1, while that for the APCVD-grown h-BN in (b) is boron-rich.

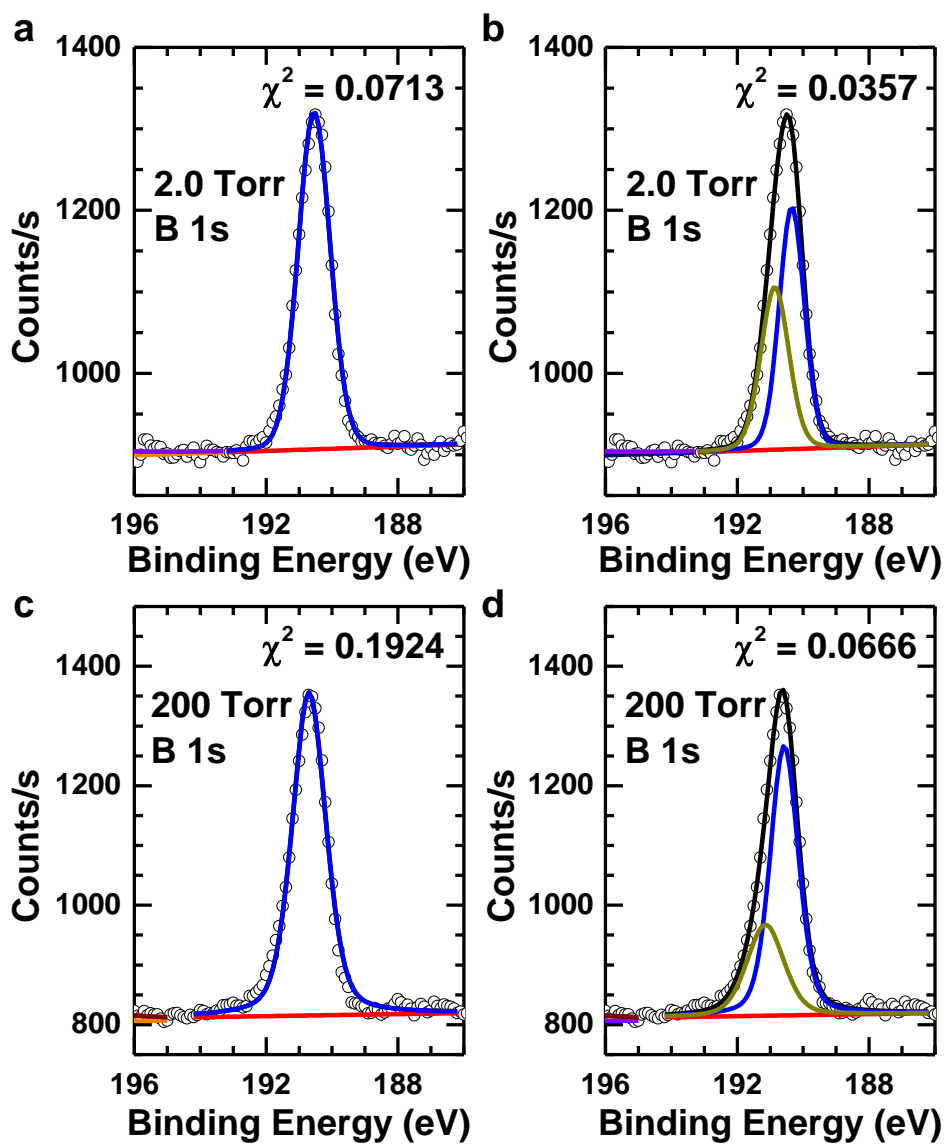

**Figure S7. Fitting comparisons for XPS B 1s core level data. (a)** Core level data for CVD h-BN grown at 2.0 Torr, fitted with one Gaussian-Lorentzian (GL) lineshape about the main $sp^2$ B 1s peak. We hold that accurate fits are achieved when the residual value, given by $\chi^2$, is less than 0.1, as is the case for (a). **(b)** The same core level data as (a), fitted with two GL lineshapes under the main $sp^2$ B 1s peak. A sub-peak at higher binding energy (BE) under the main $sp^2$ B 1s peak can correspond to $sp^3$-like B domains within the film,[8] similar to graphene C 1s XPS spectra. The $\chi^2$ value for (b) is less than (a), but since both are less than 0.1, the second GL sub-peak at higher BE is redundant. Thus, the number of $sp^3$ domains in the 2 Torr h-BN growth is minimal. **(c)** Core level data for CVD h-BN grown at 200 Torr. A single GL fit about the $sp^2$ B 1s peak gives a high $\chi^2$ value, implying that some $sp^3$ domains might be present. **(d)** The same core level data as (c), fit with two GL lineshapes under the main $sp^2$ B peak. Compared to (c), the lower $\chi^2$ value asserts that two sub-peaks are necessary. Hence, the 200 Torr h-BN growth has some $sp^3$ structure. Nevertheless, the absence of the c-BN bulk plasmon peak[8] both show that the $sp^3$ domains are not in a c-BN configuration. The $sp^3$ domains likely originate from partially dehydrogenated $H_3N$–$BH_3$ species and the poorer crystalline nature of the films grown at higher pressures.

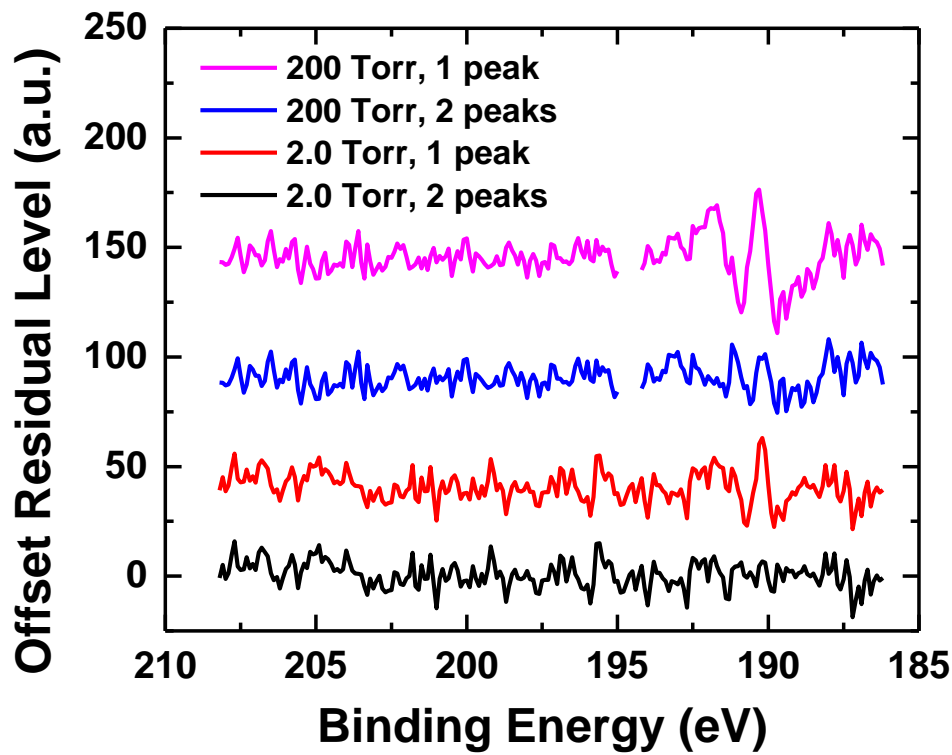

**Figure S8. Residuals from 1 and 2 peak fitting of XPS B 1s core level data for growths shown in prior figure.** Comparison of the residuals from fitting the B 1s core level data for the LPCVD (2.0 Torr) h-BN growth shows only a small difference between using 1 GL lineshape and using 2 GL lineshapes, indicating that 1 GL lineshape is sufficient. The difference between using 1 GL lineshape and 2 GL lineshapes is more obvious for the higher pressure growth (200 Torr). In turn, 1 GL lineshape is insufficient to fit the B 1s core level for the 200 Torr growth.

| Identifier | Pressure (Torr) | AFM Thickness (nm) | Intensity, Cu 2p$^{3/2}$ (cps) |
|---|---|---|---|
| h-BN #3 | 2.0 (high mass) | 3.2 ± 1.4 | – |
| h-BN #10 | 2.0 | 1.0 ± 0.3 | – |
| h-BN #11 | 2.0 | – | 51897.9 |
| h-BN #14 | 2.0 (different foil) | – | 40690.9 |
| h-BN #17 | 20 | 3.4 ± 0.6 | 39312.3 |
| h-BN #13 | 200 | 10.1 ± 0.9 | 39651.2 |
| h-BN #12 | 760 | 6.7 ± 1.3 | 34788.4 |

**Table S2. XPS Cu 2p$^{3/2}$ peak intensity versus growth pressure.** With a relatively homogeneous h-BN overlayer on the Cu, the Cu 2p$^{3/2}$ peak intensity decreases with increasing thickness. Thus, the grown films are thickest at APCVD, which is qualitatively supported by the AFM step height data. Using the step height data for h-BN grown at 200 Torr (10.1 ± 0.9 nm), the Cu 2p$^{3/2}$ sub-peak intensities, and the following equation for photoelectron attenuation ( $I_{hBN} = I_{Cu} \exp\left(-t/\lambda \cos\theta\right)$ ), we can estimate the additional overlayer thickness for the 760 Torr growth. We find the additional thickness to be 7.7 ± 0.7 nm, resulting in a layer thickness of 17.8 ± 1.1 nm after propagating uncertainty. Here, the takeoff angle $\theta = 54.5°$ and the average thickness is integrated versus the effective attenuation length $\lambda$.

h-BN #11 ($P_{TOT}$ = 2.0 Torr): **Stoichiometry BN$_x$, x = 0.95 (survey BN$_x$, x = 1.22)**

| Peak | Binding Energy (eV) | FWHM Γ (eV) | Gaussian-Lorentzian Mixing (0%=Gaussian, 100%=Lorentz) | Area (cps·eV) |
|---|---|---|---|---|
| sp$^2$ B | 190.5 | 1.14 | 12 | 532.9 |
| sp$^3$ B | – | – | – | – |
| sp$^2$ N | 398.1 | 1.15 | 0 | 1866.7 |

**Table S3. XPS sub-peak parameters for h-BN grown by LPCVD at 2.0 Torr growth pressure.** More insulating films have a Gaussian-Lorentzian mixing of 0.

h-BN #14 ($P_{TOT}$ = 2.0 Torr, on Basic Copper): **Stoichiometry BN$_x$, x = 1.04 (survey BN$_x$, x = 1.03)**

| Peak | Binding Energy (eV) | FWHM Γ (eV) | Gaussian-Lorentzian Mixing (0%=Gaussian, 100%=Lorentz) | Area (cps·eV) |
|---|---|---|---|---|
| sp$^2$ B | 190.7 | 1.22 | 1 | 368.4 |
| sp$^3$ B | – | – | – | – |
| sp$^2$ N | 398.1 | 1.20 | 7 | 1400.6 |

**Table S4. XPS sub-peak parameters for h-BN grown by LPCVD at 2.0 Torr growth pressure on a different Cu substrate.**[9] More insulating films have a Gaussian-Lorentzian mixing of 0.

h-BN #17 ($P_{\text{TOT}}$ = 20 Torr): **Stoichiometry BN$_x$, x = 0.89 (survey BN$_x$, x = 1.07)**

| Peak | Binding Energy (eV) | FWHM Γ (eV) | Gaussian-Lorentzian Mixing (0%=Gaussian, 100%=Lorentz) | Area (cps·eV) |
|------|---------------------|-------------|---------------------------------------------------------|---------------|
| sp$^2$ B | 190.6 | 1.15 | 30 | 532.5 |
| sp$^3$ B | 191.3 | 1.86 | 0 | 227.3 |
| sp$^2$ N | 398.2 | 1.31 | 10 | 2498.2 |

**Table S5. XPS sub-peak parameters for h-BN grown by CVD at 20 Torr growth pressure.** More insulating films have a Gaussian-Lorentzian mixing of 0. The onset of the sp$^3$ B sub-peak[1] confirms a larger polymeric aminoborane[10] contribution in the CVD film.

h-BN #13 ($P_{\text{TOT}}$ = 200 Torr): **Stoichiometry BN$_x$, x = 0.81 (survey BN$_x$, x = 0.92)**

| Peak | Binding Energy (eV) | FWHM Γ (eV) | Gaussian-Lorentzian Mixing (0%=Gaussian, 100%=Lorentz) | Area (cps·eV) |
|------|---------------------|-------------|---------------------------------------------------------|---------------|
| sp$^2$ B | 190.6 | 1.00 | 23 | 490.1 |
| sp$^3$ B | 191.1 | 1.40 | 34 | 309.7 |
| sp$^2$ N | 398.2 | 1.19 | 11 | 2712.1 |

**Table S6. XPS sub-peak parameters for h-BN grown by CVD at 200 Torr growth pressure.** More insulating films have a Gaussian-Lorentzian mixing of 0. The onset of the sp$^3$ B sub-peak[1] confirms a larger polymeric aminoborane[10] contribution in the CVD film.

h-BN #12 ($P_{\text{TOT}}$ = 760 Torr): **Stoichiometry BN$_x$, x = 1.03 (survey BN$_x$, x = 0.85)**

| Peak | Binding Energy (eV) | FWHM Γ (eV) | Gaussian-Lorentzian Mixing (0%=Gaussian, 100%=Lorentz) | Area (cps·eV) |
|------|---------------------|-------------|---------------------------------------------------------|---------------|
| sp$^2$ B | 190.6 | 1.16 | 30 | 459.17 |
| sp$^3$ B | 191.5 | 2.01 | 30 | 93.7 |
| sp$^2$ N | 398.2 | 1.2 | 55.96 | 2079.85 |
| organic N | 400.3 | 0.88 | 55.96 | 29.83 |

**Table S7. XPS sub-peak parameters for h-BN grown by APCVD at 760 Torr growth pressure.** More insulating films have a Gaussian-Lorentzian mixing of 0. The onset of the sp$^3$ B sub-peak[1] confirms a largercontribution of partially decomposed NH$_3$-BH$_3$ precursor byproducts[10] in the CVD film.

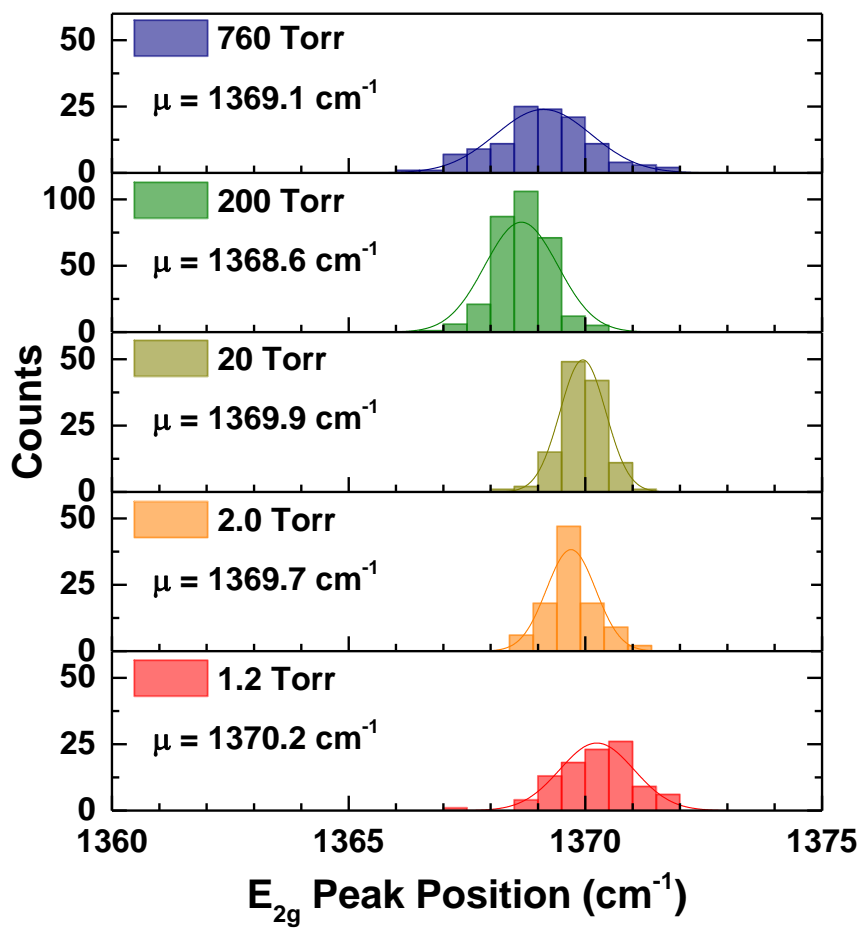

**Figure S9. Histograms of E$_{2g}$ peak position for Raman spectra for different h-BN growth pressures.** The $E_{2g}$ average peak position decreases as growth pressure increases. This indicates a thicker h-BN films[11] as the growth pressure increases. The decrease in average peak position is less than expected,[11] especially for a 10 nm thick h-BN film, as in the 200 Torr case.

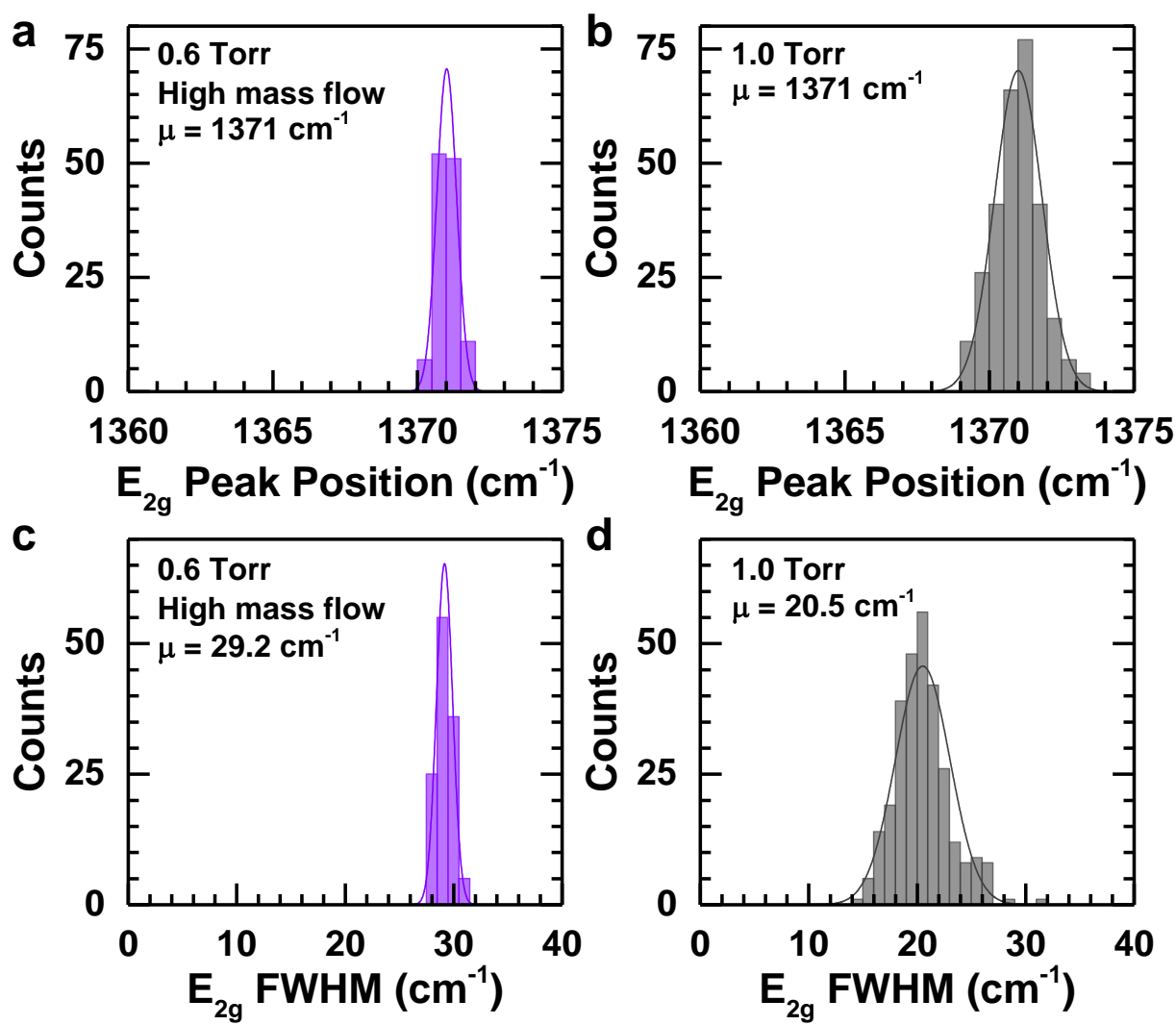

**Figure S10. Raman histograms for h-BN grown with differing precursor amounts.** Raman $E_{2g}$ band position for LPCVD h-BN growth with high precursor flux condition (**a**) and LPCVD h-BN growth with normal precursor flux condition (**b**). Due to a precursor problem during the growth in (a), there was a brief spike in the pressure when opening the valve to the precursor ampoule from the 0.6 Torr background to 10 Torr. Such conditions greatly change the ratio of precursor to $H_2$ from normal growth process. Similar to the comparison between LPCVD and high precursor flux LPCVD h-BN shown in Figure 5, the h-BN from (**a**) was thicker than that from (**b**), despite similar growth pressures. Despite the different film thicknesses between the two growths, they share the same average $E_{2g}$ band position at ~1371 cm$^{-1}$. Comparing the full-width at half-maximum (FWHM) of the $E_{2g}$ Raman mode for the two cases (**c,d**) shows that the high precursor flux LPCVD h-BN growth has a significantly higher FWHM compared to the LPCVD growth with normal precursor flux. The larger FWHM value indicates that the high precursor flux LPCVD h-BN growth is nanocrystalline,[3] despite the low Ar/$H_2$ background growth pressure used. Thus, the larger ratios of precursor to Ar/$H_2$ carrier gas create h-BN with smaller crystallite size. The h-BN film from (**a, c**) was thick, with a step height of ~46 nm. This implies that in the high mass flow regime the surface catalysis reaction rate is dominated by the mass transport reaction rate. Ultimately, the film's thickness is determined by how much $H_3N$–$BH_3$ can diffuse through a gas boundary layer. The high level of sensitivity for the h-BN growth reaction rates is in striking contrast to graphene.[12]

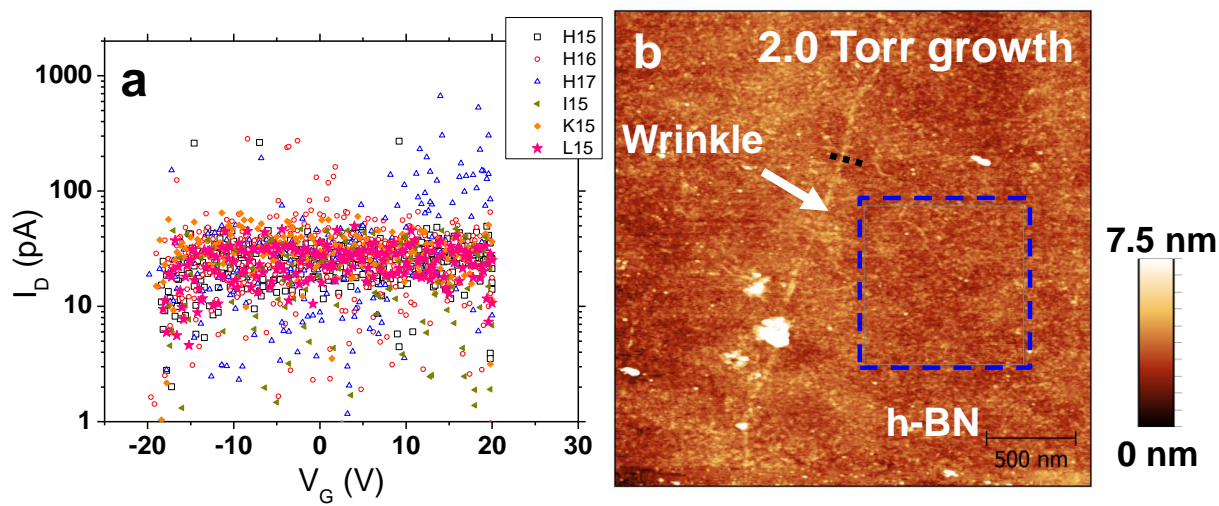

**Figure S11. Two point probe device transport on h-BN grown at 2.0 Torr.** (**a**) $I_D$–$V_G$ measurements on h-BN between two Au/Ti contacts. Current $I_D$ is comparable to the noise floor for the probe station. This shows that the h-BN is not conducting. (**b**) Atomic force microscopy (AFM) image of a representative device from (a). RMS roughness is 0.51 nm within the blue box, and the image shows a wrinkle introduced from transfer. Despite these features, the h-BN is relatively smooth. This h-BN for this device is from the same growth as that shown in Figure 2b, after annealing in air at 500 °C for 1 hr. This h-BN growth is ~3.2 nm thick.